\pdfoutput=1
\documentclass[ALICE,manyauthors]{cernphprep}

\usepackage[comma,square,numbers,sort&compress]{natbib}

\usepackage[T1]{fontenc} % if needed
\usepackage[utf8]{inputenc}
\usepackage[english]{babel}

\usepackage{amssymb}
\usepackage{hyperref}
\usepackage{lineno}
\usepackage{subfigure}
\usepackage[amssymb]{SIunits}
\usepackage{booktabs}
\usepackage{multirow}

\newcommand{\pp}{\ensuremath{\mbox{p\kern-0.05em p}}}
\newcommand{\ppbar}{\ensuremath{\mathrm{p\kern-0.05em \bar{p}}}}
\newcommand{\pPb}{\ensuremath{\mbox{p--Pb}}}
\newcommand{\PbPb}{\ensuremath{\mbox{Pb--Pb}}}
\newcommand{\sqrtS}{\ensuremath{\sqrt{s}}}
\newcommand{\sqrtSnn}{\ensuremath{\sqrt{s_{\mathrm{NN}}}}}
\newcommand{\sqrtSE}[2][TeV]{\ensuremath{\sqrtS = #2~\mathrm{#1}}}
\newcommand{\sqrtSnnE}[2][TeV]{\ensuremath{\sqrtSnn = #2~\mathrm{#1}}}
\newcommand{\GeVc}{\ensuremath{\mathrm{GeV}\kern-0.05em/\kern-0.02em c}}
\def\pt#1{\ensuremath{p_{\rm T#1}}} 
\def\jt#1{\ensuremath{j_{\rm T#1}}}
\def\vjt#1{\ensuremath{\vec{j}_{\rm T#1}}}

\newcommand{\xlong} {\ensuremath{x_{\parallel}}}
\def\mean#1{\left<#1\right>}
\def\rms#1{\ensuremath{\sqrt{\left<#1^2\right>}}}
\def\fig#1{Fig.~\ref{#1}}
\def\eq#1{Eq.~\eqref{#1}}

\begin{document}

%%%%%%%%%%%%%%%  Title page %%%%%%%%%%%%%%%%%%%%%%%%
\begin{titlepage}
\PHyear{2018}
\PHnumber{303}      % required, will be obtained from PH
\PHdate{4 November}  % required, will be obtained from PH
%

%%% Put your own title + short title here:
\title{Jet fragmentation transverse momentum measurements from di-hadron correlations in $\sqrtSE{7}$ $\pp$ and $\sqrtSnnE{5.02}$ $\pPb$ collisions}
\ShortTitle{Jet fragmentation transverse momentum}   % appears on right page headers

%%% Do not change the next lines
\Collaboration{ALICE Collaboration\thanks{See Appendix~\ref{app:collab} for the list of collaboration members}}
\ShortAuthor{ALICE Collaboration} % appears on left page headers, do not change

\begin{abstract}
The transverse structure of jets was studied via jet fragmentation transverse momentum ($\jt{}$) distributions, obtained using two-particle correlations in proton-proton and proton-lead collisions, measured with the ALICE experiment at the LHC. The highest transverse momentum particle in each event is used as the trigger particle and the region $3 < \pt{t} < 15\,\GeVc$ is explored in this study. The measured distributions show a clear narrow Gaussian component and a wide non-Gaussian one. Based on \textsc{Pythia} simulations, the narrow component can be related to non-perturbative hadronization and the wide component to quantum chromodynamical splitting. The width of the narrow component shows a weak dependence on the transverse momentum of the trigger particle, in agreement with the expectation of universality of the hadronization process. On the other hand, the width of the wide component shows a rising trend suggesting increased branching for higher transverse momentum. The results obtained in $\pp$ collisions at $\sqrtSE{7}$ and in $\pPb$ collisions at $\sqrtSnnE{5.02}$ are compatible within uncertainties and hence no significant cold nuclear matter effects are observed. The results are compared to previous measurements from CCOR and PHENIX as well as to \textsc{Pythia}~8 and Herwig~7 simulations.
\end{abstract}
\end{titlepage}
\setcounter{page}{2}

%%%%% Body of the article
% !TEX root = jtPaperPreview.tex
%\linenumbers

\section{Introduction}
\label{sec:introduction}

Jets are collimated sprays of hadrons originating from the fragmentation of hard partons produced in high-energy particle collisions. Studying the jet fragmentation can provide information about QCD color coherence phenomena, such as angular ordering~\cite{basicsofpqcd}, and constrain hadronization models~\cite{introPythia81,herwigManual,herwig7releaseNote}. The transverse fragmentation of partons is often studied using the jet fragmentation transverse momentum, $\jt{}$, that describes the momentum component of particles produced in the fragmentation perpendicular to the momentum vector of the hard parton initiating the fragmentation. Previously, $\jt{}$ has been studied using two-particle correlations by the CCOR collaboration at ISR with $\pp$ collisions at center-of-mass energy $\sqrtS = 31,\;45$ and $63~\mathrm{GeV}$~\cite{firstjtmeasurement} and the PHENIX collaboration at RHIC with $\pp$ collisions at $\sqrtSE[GeV]{200}$~\cite{PHENIXjets} and d--Au collisions at center-of-mass energy per nucleon pair $\sqrtSnnE[GeV]{200}$~\cite{phenixJtPAu}. Jet measurements to study $\jt{}$ have been done by the CDF collaboration at the Tevatron with $\ppbar$ collisions at $\sqrtSE{1.96}$~\cite{cdfpaper} and the ATLAS collaboration at the LHC with $\PbPb$ collisions at $\sqrtSnnE{2.76}$~\cite{atlaksenJetit}.

Jet fragmentation in QCD consists of two different steps~\cite{eventGenerators}. After the hard scattering, partons go through a QCD induced showering step, where gluons are emitted and the high virtuality  of the parton is reduced. Since the transverse momentum scale ($Q^{2}$) is large during the showering, perturbative QCD calculations can be applied. When $Q^{2}$ becomes of the order of $\Lambda_{\mathrm{QCD}}$, partons hadronize into final-state particles through a non-perturbative process. Two distinct components, related to the showering and hadronization phases, can be identified from the measured $\jt{}$~distributions.

The presence of a heavy nucleus as in p--A collisions might alter the fragmentation process. One possible mechanism for this is initial or final-state scattering of partons inside the nucleus. This is expected to lead to a broadening of jets, since the scattered partons are likely to deviate from their original direction~\cite{jetBroadeningPpb1}. Also the nuclear parton distribution functions can change the relative contributions of quarks and gluons compared to free nucleons, for example via gluon saturation and shadowing effects~\cite{introCgc,eps09}. Understanding the implications of these cold nuclear matter effects will provide an important baseline for similar measurements in heavy-ion collisions.

In this paper, the $\jt{}$ distributions are studied using two-particle correlations, measured by the ALICE detector in $\sqrtSE{7}$ $\pp$ and $\sqrtSnnE{5.02}$ $\pPb$ collisions. The correlation approach is chosen as opposed to full jet reconstruction based on the discussion in Ref.~\cite{thorstenBiasProceedings,Renk:2011wp}, where it is argued that two-particle correlations are more sensitive to the soft and non-perturtabive parts of the jet fragmentation. This is important for the separation of the two $\jt{}$ components and in searching for cold nuclear matter effects that are expected to play a larger role at lower momenta.

This paper is structured as follows. The event and track selection together with the used data samples are described in Section~\ref{sec:experimentaldetails}. The analysis details are discussed in Section~\ref{sec:methods}, followed by the systematic uncertainty analysis in Section~\ref{sec:systematicerrors}. The obtained results are shown in Section~\ref{sec:results} and the observations are summarized in Section~\ref{sec:conclusions}.

\section{Experimental setup and data samples}
\label{sec:experimentaldetails}

This analysis uses two different datasets. The $\sqrtSE{7}$ $\pp$ ($3.0 \cdot 10^{8}$ events, integrated luminosity $\mathcal{L}_{\mathrm{int}} = \unit{4.8}{nb^{-1}}$) collisions were recorded in 2010 and the $\sqrtSnnE{5.02}$ $\pPb$ ($1.3 \cdot 10^{8}$ events, $\mathcal{L}_{\mathrm{int}} = \unit{620}{nb^{-1}}$) collisions were recorded in 2013 by the ALICE detector~\cite{aliceDetector}. The details of the performance of the ALICE detector during LHC Run~1 (2009-2013) are presented in Ref.~\cite{alicePerformance}.

The charged particle tracks used in this analysis are reconstructed using the Inner Tracking System (ITS)~\cite{aliceITS} and the Time Projection Chamber (TPC)~\cite{aliceTPC}. The tracking detectors are located inside a large solenoidal magnet which provides a homogeneous magnetic field of $\unit{0.5}{T}$. They are used to reconstruct the tracks within a pseudorapidity range of $|\eta| < 0.9$ over the full azimuth. The ITS consists of six layers of silicon detectors: the two innermost layers are the Silicon Pixel Detector (SPD), the two middle layers are the Silicon Drift Detector (SDD) and the two outermost layers are the Silicon Strip Detector (SSD). The TPC is a gas-filled detector capable of providing three-dimensional tracking information over a large volume. Combining information from the ITS and the TPC, the momenta of charged particles from $0.15$ to $100~\GeVc$ can be determined with a resolution ranging from $1$ to $10\,\%$. For tracks without the ITS information, the momentum resolution is comparable to that of ITS+TPC tracks below transverse momentum $\pt{} = 10~\GeVc$, but for higher momenta the resolution reaches $20\,\%$ at $\pt{} = 50~\GeVc$~\cite{alicePerformance,aliceBackgroundFluctuation}. Charged particle tracks with $\pt{} > 0.3~\GeVc$ in the region $|\eta| < 0.8$ are selected for the analysis.
Events are triggered based on the information of the V0 detector~\cite{forwarddetectorsTdr} together with the SPD. The V0 detector consists of two scintillator stations, one on each side of the interaction point, covering $-3.7 < \eta < -1.7$ (V0C) and $2.8 < \eta < 5.1$ (V0A). For the 2010 $\pp$ collisions, the minimum bias (MB) triggered events are required to have at least one hit from a charged particle traversing the SPD or either side of the V0. The pseudorapidity coverage of the SPD is $|\eta| < 2$ in the first layer and $|\eta| < 1.5$ in the second layer. Combining this with the acceptance of the V0, the particles are detected in the range $-3.7 < \eta < 5.1$. The minimum bias trigger definition for the 2013 $\pPb$ collisions is slightly different. Events are required to have signals in both V0A and V0C. This condition is also used later offline to reduce the contamination of the data sample from beam-gas events by using the timing difference of the signal between the two stations~\cite{alicePerformance}.

For the $\pp$ collisions, similar track cuts as in Ref.~\cite{ALICE:2011ac} are used: at least two hits in the ITS are required, one of which needs to be in the three innermost layers, and 70 hits out of 159 are required in the TPC. In addition, the distance of the closest approach (DCA) of the track to the primary vertex is required to be smaller than $\unit{2}{cm}$ in the beam direction. In the transverse direction, a $\pt{}$ dependent cut DCA $< \unit{0.0105}{cm} + \unit{0.035}{cm} \cdot \pt{}^{-1.1}$ is used, where $\pt{}$ is measured in units of $\GeVc$. These track cuts are tuned to minimize the contamination from secondary particles.

For the $\pPb$ collisions the tracks are selected following the so called hybrid approach, which is described in detail in Ref.~\cite{hybridExplanation}. This approach differs from the one presented above in the selection of ITS tracks. The tracks with at least one hit in the SPD and at least two hits in the whole ITS are always accepted. In addition, tracks with fewer than two hits in the ITS or no hits in the SPD are accepted, but only if an additional vertex constraint is fulfilled. The DCA cuts are also looser: smaller than $\unit{3.2}{cm}$ in the beam direction and smaller than $\unit{2.4}{cm}$ in the transverse direction. With this track selection, the azimuthal angle ($\varphi$) distribution is as uniform as possible, because it is not affected by dead regions in SPD. This is important for a two-particle correlation analysis. The momentum resolutions of the two classes of particles are comparable up to $\pt{} \approx 10\;\GeVc$, but after that, tracks without ITS requirements have a worse resolution~\cite{alicePerformance,aliceBackgroundFluctuation}.

\section{Analysis method}
\label{sec:methods}

The analysis is performed by measuring two-particle correlation functions. In each event, the trigger particle is chosen to be the charged particle with the highest reconstructed $\pt{}$ inside the acceptance region, called the leading particle. For the momentum range studied in the analysis, simulation studies show that the direction of the leading particle can be assumed in good approximation that one of the jet axis, which is the axis defined by the momentum vector of the hard parton initiating the jet fragmentation. The associated particles close in the phase-space to the leading one are then interpreted as jet fragments.

\begin{figure}
    \begin{center}
      \includegraphics[width = 0.30\textwidth]{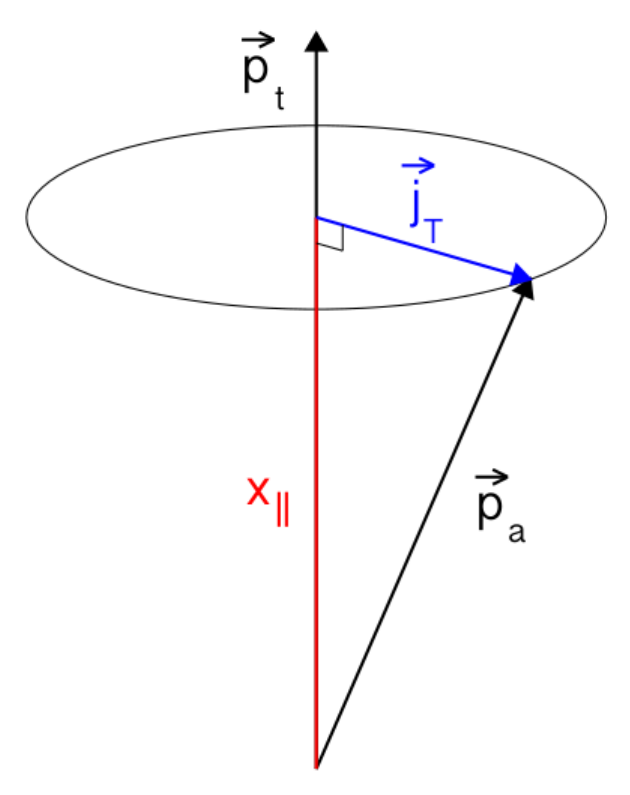}
    \end{center}
    \caption{(Color online). Illustration of $\vjt{}$ and $\xlong$. The jet fragmentation transverse momentum, $\vjt{}$, is defined as the transverse momentum component of the associated particle momentum, $\vec{p}_{\mathrm{a}}$, with respect to the trigger particle momentum, $\vec{p}_{\mathrm{t}}$. The fragmentation variable $\xlong$ is the projection of $\vec{p}_{\mathrm{a}}$ to $\vec{p}_{\mathrm{t}}$ divided by $p_{\mathrm{t}}$.}
    \label{fig:jtdefinition}
  \end{figure}

The jet fragmentation transverse momentum, $\jt{}$, is defined as the component of the associated particle momentum, $\vec{p}_{\mathrm{a}}$, transverse to the trigger particle momentum, $\vec{p}_{\mathrm{t}}$. The resulting $\vjt{}$ is illustrated in~\fig{fig:jtdefinition}. The length of the $\vjt{}$ vector is
  \begin{equation}
    \jt{} = \frac{|\vec{p}_{\mathrm{t}} \times \vec{p}_{\mathrm{a}}|}{|\vec{p}_{\mathrm{t}}|} \,.
  \label{eq:jtdefinition}
  \end{equation}
It is commonly interpreted as a transverse kick with respect to the initial hard parton momentum that is given to a fragmenting particle during the fragmentation process. In other words, $\jt{}$ measures the momentum spread of the jet fragments around the jet axis. In the analysis, results are presented in bins of the fragmentation variable $\xlong$ which is defined as the projection of the momentum of the associated to the trigger particle one, divided by the momentum of the trigger particle: 
  \begin{equation}
    \xlong = \frac{\vec{p}_{\mathrm{t}} \cdot \vec{p}_{\mathrm{a}}}{\vec{p}_{\mathrm{t}}^{2}} \,.
  \label{eq:xedefinition}
  \end{equation}
This is also illustrated in~\fig{fig:jtdefinition}. Because $\xlong$ is defined as a fraction of the trigger particle momentum, it is intuitive to define a three-dimensional near side with respect to the axis defined by the trigger momentum. The associated particle is defined to be in the near side if it is in the same hemisphere as the trigger particle:
  \begin{equation}
    \vec{p}_{\mathrm{t}} \cdot \vec{p}_{\mathrm{a}} > 0 \,.
    \label{eq:3Dnearside}
  \end{equation}
The results have been binned in $\xlong$ rather than associated particle transverse momentum ($\pt{a}$) because the definition of $\jt{}$ (\eq{eq:jtdefinition}) has an explicit $\pt{a}$ dependence. Bins in $\pt{a}$ would bias the results since pairs with larger $\jt{}$ are more likely to be in bins of larger $\pt{a}$. In the case of $\xlong$ this bias is not present, since $\xlong$ and $\jt{}$ measure momentum components along perpendicular axes. Another advantage for using $\xlong$ is that the relative $\pt{}$ of the associate particles with respect to trigger $\pt{}$ ($\pt{t}$) stays the same in different $\pt{t}$ bins. It was verified with a \textsc{Pythia}~8~\cite{introPythia81,introPythia82} Monash tune simulation that the average fraction of the leading parton momentum taken by the leading particle ($\mean{z_{\mathrm{t}}}$) varies less than 0.05 units inside the used $\xlong$ bins $0.2 < \xlong < 0.4$, $0.4 < \xlong < 0.6$, and $0.6 < \xlong < 1.0$, with lower $\pt{t}$ bins having slightly larger $\mean{z_{\mathrm{t}}}$ than higher bins.
  
The extracted $\jt{}$ distribution is of the form
\begin{equation}
  \frac{1}{N_{\mathrm{trigg}}} \frac{1}{\jt{}} \frac{\mathrm{d}N}{\mathrm{d}\jt{}} \left(\pt{t},\,\xlong,\,\jt{}\right)= C_{\mathrm{associated}}(\pt{a}) \, C_{\mathrm{Acc}}(\Delta\eta,\,\Delta\varphi) \, \frac{N_{\mathrm{pairs}}(\pt{t},\,\xlong,\,\jt{}) }{\jt{} \, N_{\mathrm{trigg}} \, \Delta \jt{}} \,,
  \label{eq:exactDistribution}
\end{equation}
where $N_{\mathrm{trigg}}$ is the number of triggers, $N_{\mathrm{pairs}}(\pt{t},\,\xlong,\,\jt{})$ is the number of trigger-associated pairs, $\Delta \jt{}$ is the bin width of the used $\jt{}$ bin, $C_{\mathrm{associated}}(\pt{a})$ is the single track efficiency correction for the associated particle and $C_{\mathrm{Acc}}(\Delta\eta,\,\Delta\varphi)$ is the pair acceptance correction. The single track efficiency correction is estimated by Monte Carlo simulations of \textsc{Pythia}~6~\cite{pythiaBig}, \textsc{Pythia}~8 or DPMJET~\cite{dpmjet} events, using GEANT3~\cite{geant} detector simulation and event reconstruction. The pair acceptance correction is the inverse of the normalized mixed event distribution sampled at the corresponding $(\Delta\eta,\,\Delta\varphi)$ value. The mixed event distribution is constructed by correlating trigger and associated particles from different events in the data sample. In the mixed event distribution, away-side particles must be included to properly correct for detector and acceptance effects.

\begin{figure}
    \begin{center}
      \includegraphics[width = 0.65\textwidth]{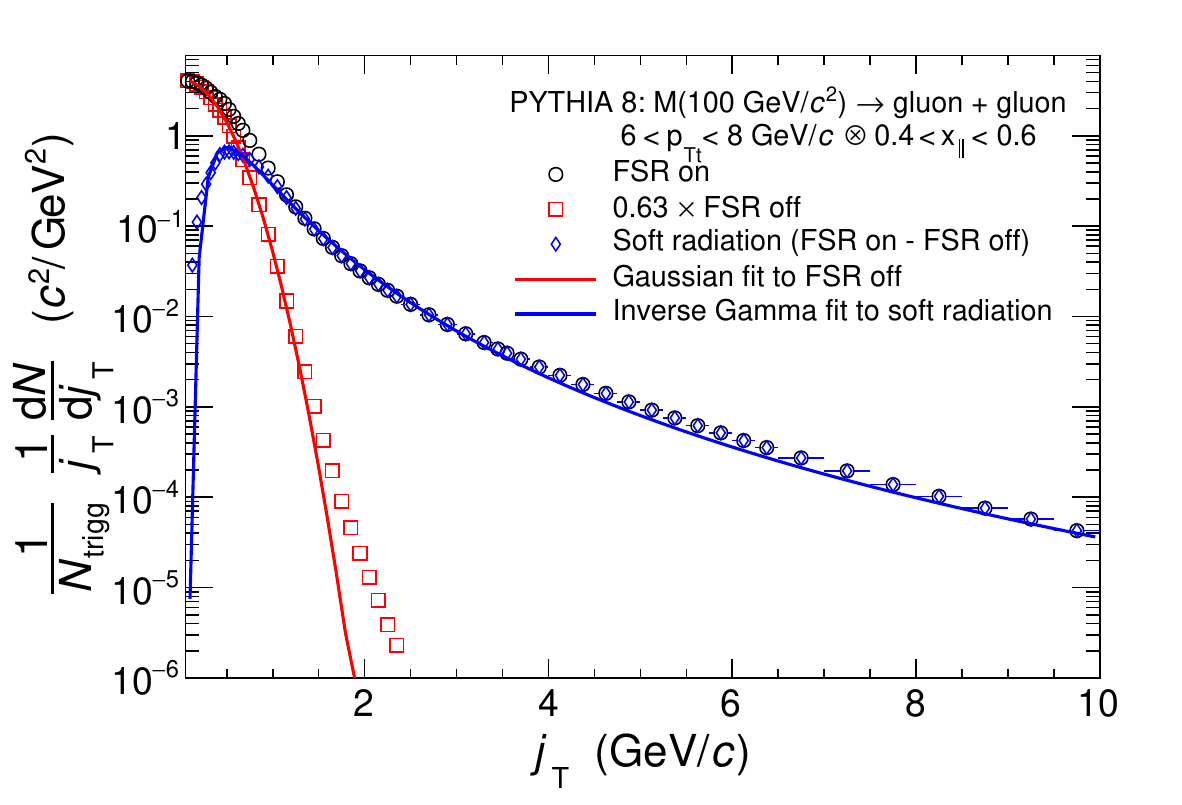}
    \end{center}
    \caption{(Color online). Results from a \textsc{Pythia}~8 study with a di-gluon initial state. The circular symbols are obtained when the final-state shower is enabled. The square symbols show the distribution without final-state showering. The diamond symbols representing soft radiation are obtained as a difference between the other two distributions. The distribution without final-state showering is fitted with a Gaussian and the soft radiation part with an inverse gamma function.}
    \label{fig:componentsFromResonance}
  \end{figure}

In this study, the $\jt{}$ distribution is determined by pairing all charged particles inside each $\xlong$ bin with the leading particle and calculating $\jt{}$ for each of these pairs in an event. After that, two distinct components are extracted from the $\jt{}$ distribution. A generator level \textsc{Pythia}~8 simulation was performed to gain support for the separation of these components. To create a clean di-jet event sample, \textsc{Pythia}~8 was initialized to produce two hard gluons with a constant invariant mass for each event. The final-state QCD shower in \textsc{Pythia}~8 is modeled as a timelike shower, as explained in Ref.~\cite{newPythiaShower}. Two simulations were studied, one where the final-state shower was present and one where it was disabled. Without the final-state shower, the hadronization of the leading parton via Lund string fragmentation~\cite{lundString} develops without a QCD showering phase preceding it. When the final-state shower is allowed, the partons go through both showering and hadronization. The results of this study are presented in \fig{fig:componentsFromResonance}. The squares show a nearly Gaussian distribution resulting from the case when the final-state shower is disabled. The circles are obtained when the final-state shower is enabled. A long tail is observed which was not seen in the case with final-state shower off. To estimate the QCD showering component, it is assumed that hadronization dominates at low $\jt{}$, and the distributions from the two simulations coincide at $\jt{} = 0$. The "hadronization only" -distribution in \fig{fig:componentsFromResonance} needs to be scaled with a factor of 0.63 for this, since without QCD splittings the partons hadronize at higher scale, producing more particles. With the subtraction of the "hadronization only" -distribution from the total one, the QCD showering part can be separated. This is represented by the diamond symbols in \fig{fig:componentsFromResonance}.
  
This study shows a possible factorization of the showering and hadronization parts of the jet fragmentation in \textsc{Pythia}~8. Based on simulations, template fit functions for hadronization and showering components have been estimated and used to extract the corresponding terms from the data. Since $\vjt{}$ is a two-dimensional vector, using two-dimensional forms for the fit functions allows to extract the final results from the functions more easily. Assuming that there is no dependence on the polar angle of the vector, the angle can be integrated out and the distributions written as a function of the length of the vector. The hadronization part can be described by a Gaussian:
\begin{equation}
f_{G}(\jt{}) = \frac{A_{2}}{A_{1}^{2}} \, e^{-\frac{\jt{}^{2}}{2 A_{1}^{2}}},
\label{eq:gauss}
\end{equation}
and the showering part by an inverse gamma function of the form:
\begin{equation}
f_{IG}(\jt{}) = \frac{A_{3} \, A_{5}^{A_{4}-1}}{\Gamma(A_{4}-1)} \, \frac{e^{-\frac{A_{5}}{\jt{}}}}{\jt{}^{A_{4}+1}}\;,
\label{eq:inverseGamma}
\end{equation}  
where $A_{1\ldots5}$ are the free fit parameters and $\Gamma$ is the gamma function. In this paper, the hadronization part will be called the narrow component and the showering part the wide component.
  
\begin{figure}
    \begin{center}
      \subfigure{ \includegraphics[width = 0.42\textwidth]{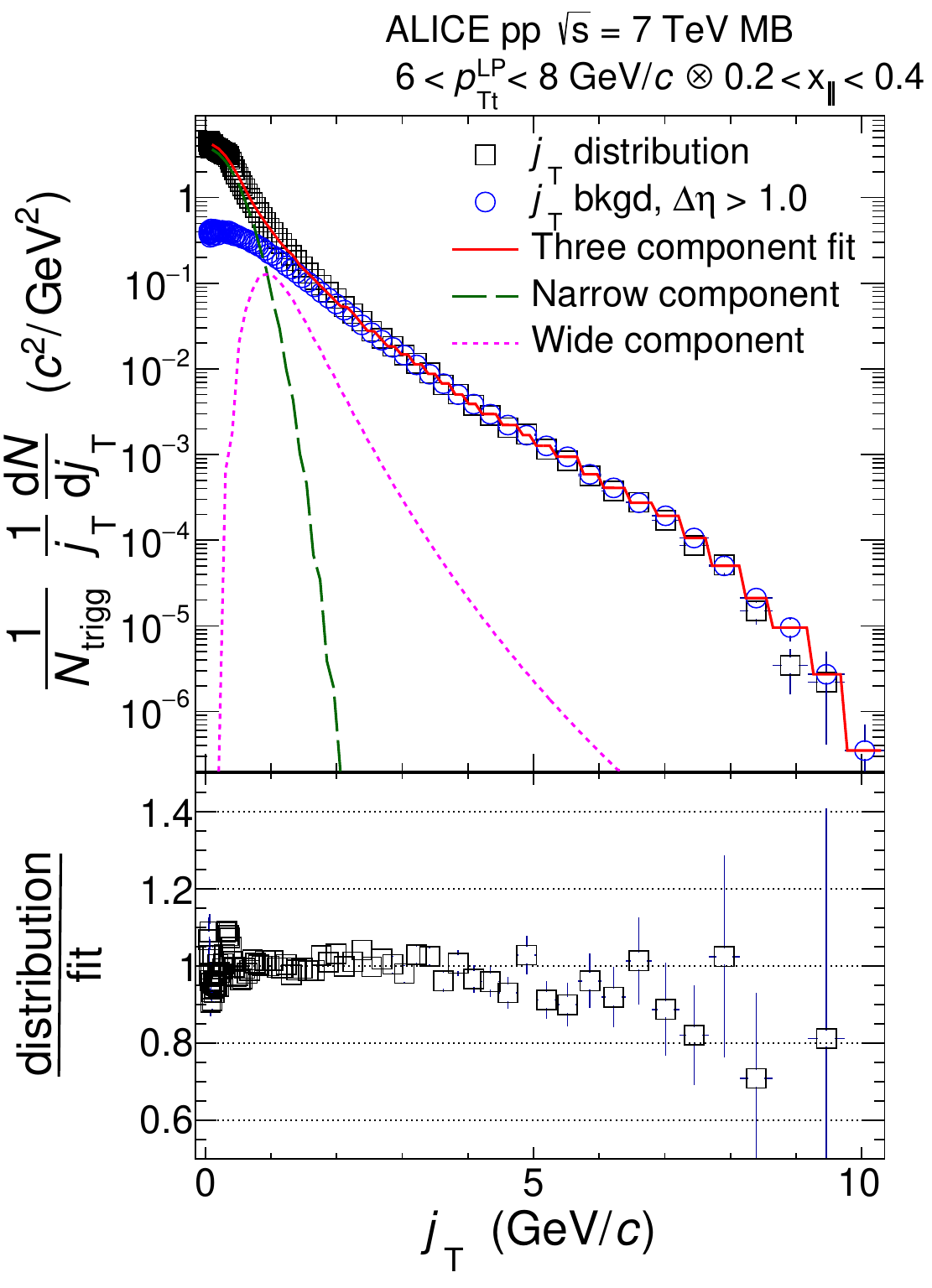}}
      \subfigure{\includegraphics[width = 0.42\textwidth]{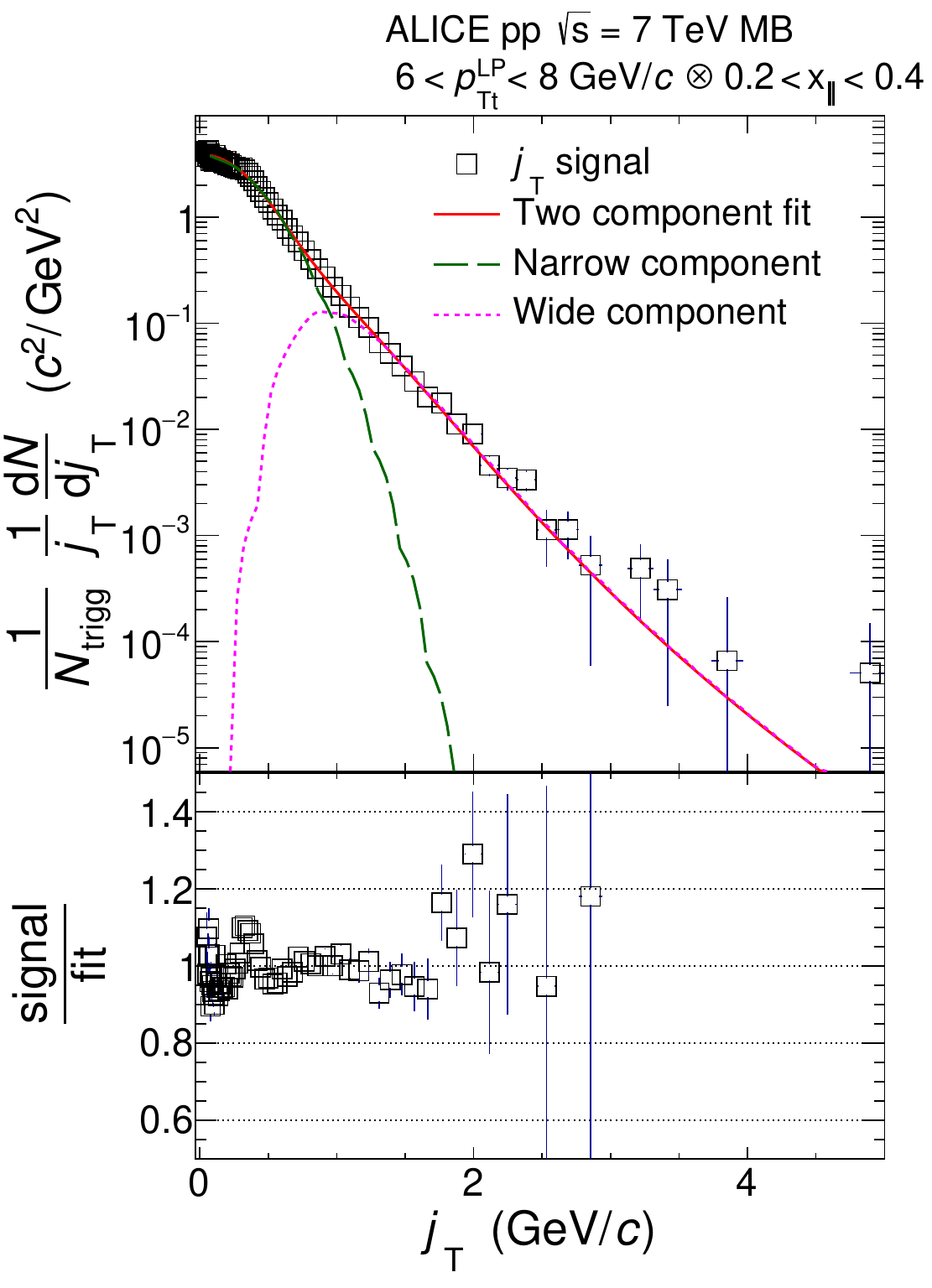}}
    \end{center}
    \caption{(Color online). \emph{Left:} Measured $\jt{}$ distribution including a three-component fit. The three components describe the background (circular symbols), hadronization (long dashed line), and showering (short dashed line). \emph{Right:} The same $\jt{}$ distribution but with background subtracted.}
    \label{fig:jtdistribution}
  \end{figure}

In the data, in addition to the signal, a background component mostly due to the underlying event is observed. Examples of measured $\jt{}$ distributions with background included and subtracted are presented in \fig{fig:jtdistribution}. An $\eta$-gap method is used to estimate the background contribution. Pairs with $|\Delta\eta| > 1.0$ are considered as background from the underlying event. The background templates for the analysis are built by randomizing the pseudorapidities for the trigger and the associated particles, following the inclusive charged particle pseudorapidity distributions. Twenty randomized pairs are generated from each background pair to improve the statistics for the background. The template histograms, generated in bins of $\pt{t}$ and $\xlong$, are then fitted to the $\jt{}$ distribution together with a sum of a Gaussian function and an inverse gamma function. It can be seen from \fig{fig:jtdistribution} that the fit is in good agreement with the data, except in the region around $\jt{} \sim 0.4~\GeVc$, where the data shows an increase with respect to the fit function. \textsc{Pythia} studies show that this structure is caused by correlations from neutral meson decays, dominated by decays of $\rho^{0}$ and $\omega$, where one of the decay daughters is the leading charged particle in the event. The effect of this structure is taken into account in the evaluation of the systematic uncertainties.

The goal of the analysis is to determine the root-mean-square $\left(\text{RMS} = \rms{\jt{}}\,\right)$ values and yields of the narrow and wide $\jt{}$ components. These are calculated from the parameters of the fit functions in equations~\eqref{eq:gauss} and~\eqref{eq:inverseGamma}.

\section{Systematic uncertainties}
\label{sec:systematicerrors}

The systematic uncertainties considered for this analysis arise from the background determination, the signal fitting procedure and the cuts used to select the tracks. The uncertainties related to the tracking are estimated from variations of the track selection cuts defined in Section~\ref{sec:experimentaldetails}. The resulting variations of the RMS and yield are below 3~\% in most cases, but effects up to 17~\% are observed for the yield of the wide component. The tracking efficiency contributes to the uncertainty of the yields only. This uncertainty is estimated from the difference between data and simulation in the TPC-ITS track matching efficiency as is previously done in Refs.~\cite{spectrumReferencePp} and~\cite{spectrumReferencepPb}.  For $\pp$ collisions this uncertainty is 5~\% and for $\pPb$ ones 4~\%. The effect due to the subleading track being reconstructed as a leading track was studied using simulations and found to be negligible due to steep slope of the trigger spectrum.

The main source of uncertainty from the background evaluation comes from the background region definition. As an alternative method to the default procedure, uncorrelated background templates are generated from particles with $R = \sqrt{\Delta\varphi^2 + \Delta\eta^2} > 1$ instead of those at large $\Delta\eta$, and pseudorapidities for the particle pairs are randomized together with azimuthal angles. The associated uncertainty is typically below 5~\%, but for the yield of the wide component the uncertainty can grow up to 46~\% in the lowest $\pt{t}$ and $\xlong$ bins where the signal to background ratio is the worst (0.84 for $\pp$ and 0.33 for $\pPb$). Changing the size of the $\eta$-gap produces small uncertainties compared to other sources, usually below 2~\%. The effect of changing the number of new pairs generated for the background from 20 to 15 or 25 was also checked, but this was found to be negligible and is not included in the total uncertainties.

The dominant source of uncertainty results from decaying neutral mesons. Even though this is a physical correlation in the $\jt{}$ distribution, it cannot be attributed to QCD showering or hadronization. The effect of the decay mesons is estimated from a variation of the fit range, excluding the region where the data shows an increase with respect to the fit function. The excluded regions are $0.25 < \jt{} < 0.45\,\GeVc$, $0.2 < \jt{} < 0.6\,\GeVc$ or $0.2 < \jt{} < 0.65\,\GeVc$ for the $\xlong$ bins  $0.2 < \xlong < 0.4$,  $0.4 < \xlong < 0.6$ and  $0.6 < \xlong < 1.0$, respectively. For the yield of the wide component the uncertainty can go up to 60~\% in the $0.4 < \xlong < 0.6$ bin at low $\pt{t}$. In most cases, this uncertainty is well below 10~\%. For the signal fit, the difference between fitting the background and the signal simultaneously and only the signal, after background subtraction, was evaluated. The uncertainty from this source was found to be typically smaller than 3~\%, which is small compared to other sources.

The different sources of systematic uncertainties were considered as uncorrelated and added in quadrature accordingly. In general, the systematic uncertainties for the wide component are larger than for the narrow component, since the signal to background ratio is significantly smaller for the wide component. Also the uncertainties for the yield are larger than for the RMS. The uncertainties are also $\pt{t}$ and $\xlong$ dependent. For different results and datasets, the total systematic uncertainties vary within the ranges summarized in Tab.~\ref{tab:uncertainties}. The smallest uncertainty of $1.6$~\% for the narrow component RMS is found for the $0.2 < \xlong < 0.4$ and highest $\pt{t}$ bins while the largest uncertainty of 73~\% for the yield  of the wide component is found from the $0.4 < \xlong < 0.6$ and lowest $\pt{t}$ bins .

\begin{table}
  \caption{Total systematic uncertainties for RMS and yield of the narrow and wide components in $\pp$ and $\pPb$ collisions. The ranges reflect $\pt{t}$ and $\xlong$ dependence of the studied observables for data.}
  \begin{center}
    \begin{tabular}{cccc}
      \toprule
                                             &       &\multicolumn{2}{c}{Total relative uncertainty} \\
                                             &       &      $\pp$        &      $\pPb$        \\
                                             \hline
        \multirow{ 2}{*}{Narrow component}   &  RMS  &  $1.6 - 6.2\,\%$  &   $1.6 - 8.5\,\%$  \\
                                             & yield &  $5.2 - 21\,\%$   &   $5.4 - 13\,\%$   \\
        \multirow{ 2}{*}{Wide component}     &  RMS  &  $1.9 - 7.4\,\%$  &   $3.0 - 14\,\%$   \\
                                             & yield &  $8.5 - 48\,\%$   &   $13  - 73\,\%$   \\
        \bottomrule
    \end{tabular}
  \end{center}
  \label{tab:uncertainties}
\end{table} 

The systematic uncertainty estimation is done also for the \textsc{Pythia} and Herwig simulations, which are compared to the data. As the same analysis method is used for simulations and data, also the same methods to estimate the systematic uncertainty can be applied. For the simulations, the uncertainty is estimated from the background determination and signal fitting.

\section{Results and discussions}
\label{sec:results}

\begin{figure}[t]
    \begin{center}
      \includegraphics[width = 0.98\textwidth]{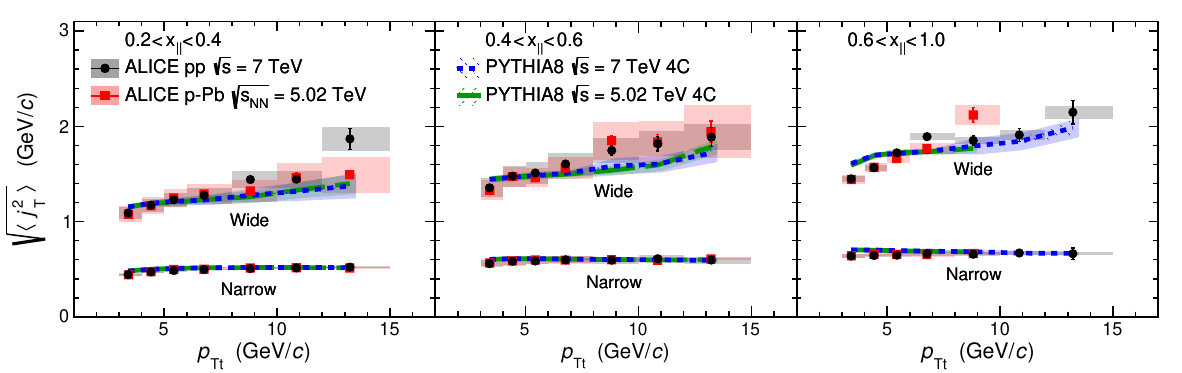}
    \end{center}
    \caption{(Color online). RMS values of the narrow and wide $\jt{}$ components. Results from $\pp$ collisions at $\sqrtSE{7}$ (circular symbols) and from $\pPb$ collisions at $\sqrtSnnE{5.02}$ (square symbols) are compared to \textsc{Pythia}~8 tune 4C simulations at $\sqrtSE{7}$ (short dashed line) and at $\sqrtSE{5.02}$ (long dashed line). Different panels correspond to different $\xlong$ bins with $0.2 < \xlong < 0.4$ on the left, $0.4 < \xlong < 0.6$ in the middle, and $0.6 < \xlong < 1.0$ on the right. The statistical errors are represented by bars and the systematic errors by boxes.}
    \label{fig:jtrms}
  \end{figure}

The per trigger yields and widths of the $\jt{}$ distributions are determined as a function of the transverse momentum of trigger particle in the range $3 < \pt{t} < 15~\GeVc$ for three $\xlong$ bins $0.2 < \xlong < 0.4$, $0.4 < \xlong < 0.6$ and $0.6 < \xlong < 1.0$. The results are obtained from the area and RMS of the fits to the narrow and wide components of the $\jt{}$ distribution. The RMS values for both components in different $\xlong$ bins from $\sqrtSE{7}$ $\pp$ and $\sqrtSnnE{5.02}$ $\pPb$ collisions are compared with \textsc{Pythia}~8 tune 4C~\cite{pythia4CTune} simulations with the same energies in \fig{fig:jtrms}. The narrow component results show only weak dependence on $\pt{t}$ in the lowest $\xlong$ bin and no dependence on $\pt{t}$ in the higher $\xlong$ bins. These behaviours is sometimes referred to as universal hadronization. There is also no difference between $\pp$ and $\pPb$ collisions. \textsc{Pythia}~8 simulations for the two energies give consistent results that are in agreement with data, within uncertainties.

Comparing the three panels in \fig{fig:jtrms}, it can be seen that $\rms{\jt{}}$ is larger in higher $\xlong$ bins for both components. Kinematically, if the opening angle is the same, larger associated momentum translates into larger $\jt{}$. Jets with larger momenta are known to be more collimated, but the net effect of these two might still increase $\mean{\jt{}}$. Also if the trigger particle is not perfectly aligned with the jet axis but there is non-negligible $\jt{}$ between these two axes, $\mean{\jt{}}$ will be widened more in the higher $\xlong$ bins.

For the wide component, it can be seen that there is a rising trend in $\pt{t}$ in both $\pp$ and $\pPb$ collisions as well as in \textsc{Pythia}~8 simulations. This can be explained by the fact that higher $\pt{}$ partons are likely to have higher virtuality, which allows for more phase space for branching thereby increasing the width of the distribution. Seeing that \textsc{Pythia}~8 simulations at $\sqrtSE{7}$ and $\sqrtSE{5.02}$ are in agreement, no difference related to the collision energy is expected in the real data either. Taking this into account, the fact that the $\pp$ and $\pPb$ agree within the uncertainties suggests that no significant cold nuclear matter effects can be observed in the kinematic range where this measurement is performed.

\begin{figure}[t]
    \begin{center}
      \includegraphics[width = 0.98\textwidth]{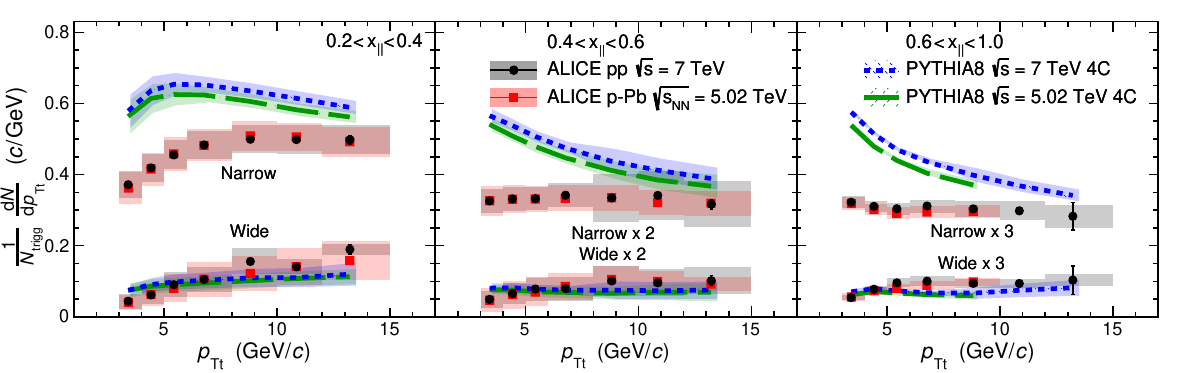}
    \end{center}
    \caption{(Color online). Yields of the narrow and wide $\jt{}$ components. Results from $\pp$ collisions at $\sqrtSE{7}$ (circular symbols) and from $\pPb$ collisions at $\sqrtSnnE{5.02}$ (square symbols) are compared to \textsc{Pythia}~8 tune 4C simulations at $\sqrtSE{7}$ (short dashed line) and at $\sqrtSE{5.02}$ (long dashed line). Different panels correspond to different $\xlong$ bins with $0.2 < \xlong < 0.4$ on the left, $0.4 < \xlong < 0.6$ in the middle, and $0.6 < \xlong < 1.0$ on the right. The statistical errors are represented by bars and the systematic errors by boxes}
    \label{fig:jtyield}
  \end{figure}

The results for the per trigger $\jt{}$ yield are presented in \fig{fig:jtyield}. The yield of the narrow component in data shows mostly no dependence on $\pt{t}$, with the exception of the lowest $\xlong$ bin where the yield rises with $\pt{t}$ for $\pt{t} < 8~\GeVc$. The trend in the \textsc{Pythia}~8 simulation is different though, the yield is decreasing as $\pt{t}$ grows. The simulation also overestimates the data for the yield of the narrow component. The discrepancy between the simulation and the data is around 50~\% in the lowest $\pt{t}$ and $\xlong$ bins. The overestimation of the yield was observed earlier in an underlying event analysis in $\pp$ collisions at $\sqrtS~=~0.9$ and $\unit{7}{TeV}$ \cite{ALICE:2011ac}.

The yield of the wide component shows a rising trend as a function of $\pt{t}$. This is expected if more splittings happen at higher $\pt{t}$, which would also explain the trend for the width. \textsc{Pythia}~8 simulations are in good agreement with the data for the yield of the wide component.
  
\begin{figure}[t]
    \begin{center}
      \includegraphics[width = 0.98\textwidth]{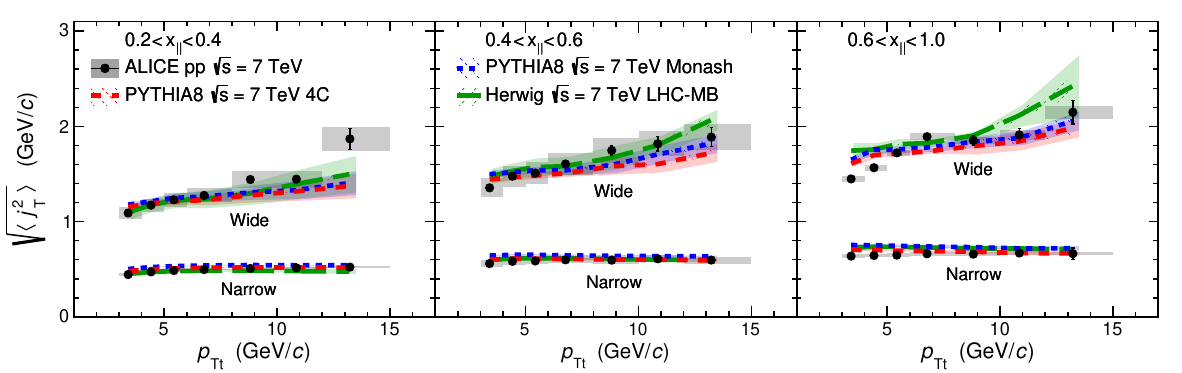}
    \end{center}
    \caption{(Color online). RMS values of the narrow and wide $\jt{}$ components for $\pp$ collisions at $\sqrtSE{7}$ (circular symbols) compared to \textsc{Pythia}~8 tunes 4C (dashed line) and Monash (short dashed line), and Herwig~7 LHC-MB tune (long dashed line) at the same energies. Different panels correspond to different $\xlong$ bins with $0.2 < \xlong < 0.4$ on the left, $0.4 < \xlong < 0.6$ in the middle, and $0.6 < \xlong < 1.0$ on the right. The statistical errors are represented by bars and the systematic errors by boxes}
    \label{fig:jtRMSmcComparison}
  \end{figure}

A comparison of the $\rms{\jt{}}$ results with different event generators and tunes is presented in \fig{fig:jtRMSmcComparison}. In this figure, the narrow and wide component $\rms{\jt{}}$ for $\sqrtSE{7}$ $\pp$ collisions are compared to \textsc{Pythia}~8 tunes 4C and Monash~\cite{pythiaMonashTune}, and to Herwig~7~\cite{herwigManual,herwig7releaseNote} tune LHC-MB. Notice that the $\pp$ data points and \textsc{Pythia}~8 tune 4C curves are the same as in \fig{fig:jtrms}. The narrow component is best described by \textsc{Pythia}~8 tune 4C. The Monash tune is approximately 10~\% above the data and Herwig~7 has a stronger $\xlong$ dependence than \textsc{Pythia}~8 or data. For the wide component, both \textsc{Pythia}~8 tunes are compatible with the data for most of the considered intervals. Herwig~7 agrees well with the data in the lowest $\xlong$ bins. All three simulation curves overestimate the RMS at low $\pt{t}$ in the $0.6 < \xlong < 1.0$ bin. At high $\pt{t}$, the central values of Herwig are larger than the data for $ \xlong > 0.4$, but the results are still consistent within the uncertainties.
  
\begin{figure}[t]
    \begin{center}
      \includegraphics[width = 0.98\textwidth]{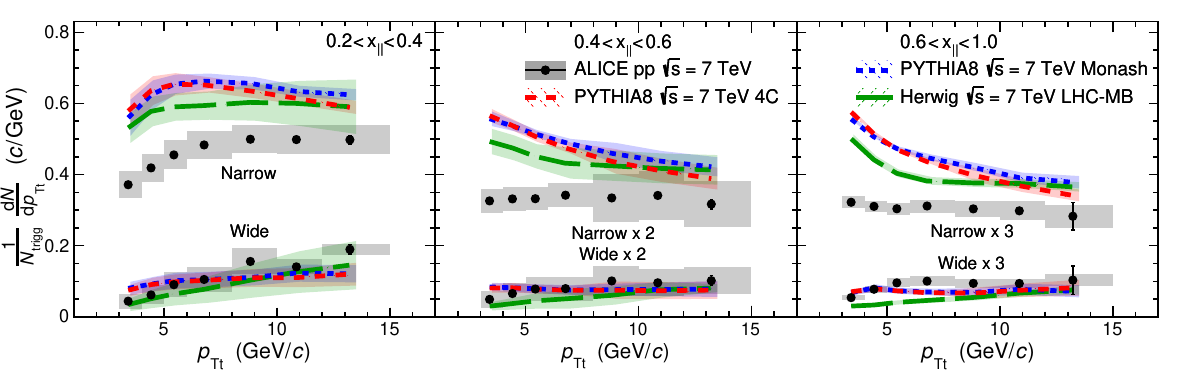}
    \end{center}
    \caption{(Color online). Yields of the narrow and wide $\jt{}$ components for $\pp$ collisions at $\sqrtSE{7}$ (circular symbols) compared to \textsc{Pythia}~8 tunes 4C (dashed line) and Monash (short dashed line), and Herwig~7 LHC-MB tune (long dashed line) at the same energies. Different panels correspond to different $\xlong$ bins with $0.2 < \xlong < 0.4$ on the left, $0.4 < \xlong < 0.6$ in the middle, and $0.6 < \xlong < 1.0$ on the right. The statistical errors are represented by bars and the systematic errors by boxes}
    \label{fig:jtyieldMcComparison}
  \end{figure}

The same \textsc{Pythia}~8 and Herwig~7 tunes are compared to the $\sqrtSE{7}$ $\pp$ yield in \fig{fig:jtyieldMcComparison}. Again, in this figure the $\pp$ and \textsc{Pythia}~8 tune 4C results are the same as in \fig{fig:jtyield}. For the narrow component, all the tunes overestimate the yield in most of the explored kinematic region. Herwig~7 shows a slightly better agreement with the data than \textsc{Pythia}~8. The relative uncertainties are quite large for the wide component and all the simulations are compatible with the data within the uncertainties in the lowest $\xlong$ bins. In the highest $\xlong$ bin, a small underestimation of the data is visible for all the simulations at mid-$\pt{t}$ and for Herwig also in the lowest $\pt{t}$ bins. 
  
\begin{figure}[t]
    \begin{center}
      \includegraphics[width = 0.5\textwidth]{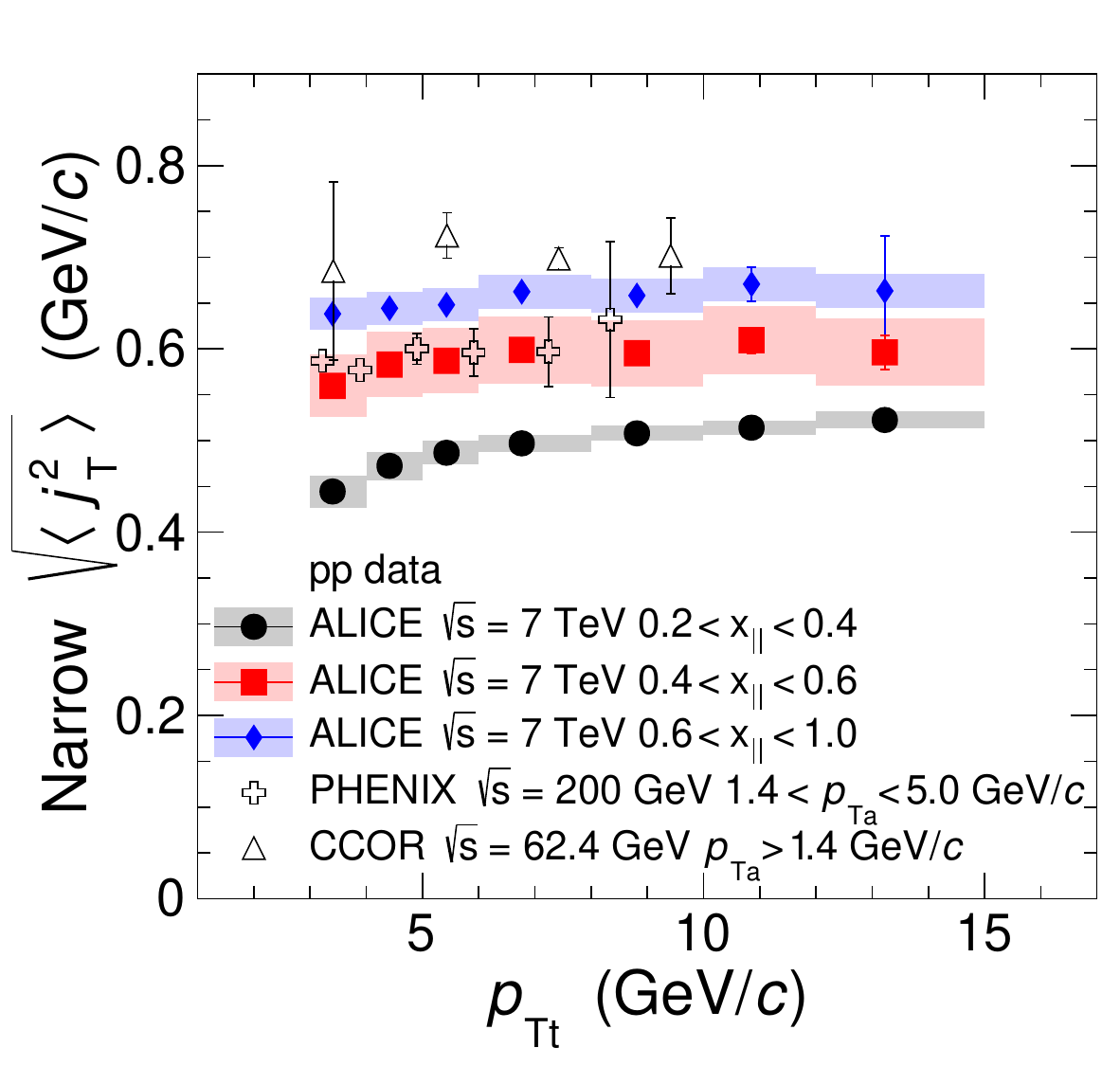}
    \end{center}
    \caption{(Color online). Narrow component RMS in different $\xlong$ bins compared with lower beam energy single component results from PHENIX~\cite{PHENIXjets} and CCOR~\cite{firstjtmeasurement}.}
    \label{fig:ancientCompare}
  \end{figure}

The narrow component $\rms{\jt{}}$ results from three $\xlong$ bins are compared to the earlier results from CCOR~\cite{firstjtmeasurement} and PHENIX~\cite{PHENIXjets} in \fig{fig:ancientCompare}. These experiments use different methods to extract $\jt{}$ from the data. In CCOR, $\jt{}$ is obtained from a fit to an away side $p_{\mathrm{out}}$ distribution, where $p_{\mathrm{out}}$ is the momentum component of a charged track going outside of the plane defined by the trigger particle and the beam axis. They use the fit function
\begin{equation}
\mean{|p_{\mathrm{out}}|}^{2} = 2 \mean{|k_{\mathrm{T}y}|}^{2} x_{\mathrm{E}}^{2} + \mean{|\jt{y}|}^{2}(1+x_{\mathrm{E}}^{2}) \,,
\label{eq:ccorjt}
\end{equation}
where $x_{\mathrm{E}} = -\vec{p}_{\mathrm{Ta}} \cdot \vec{p}_{\mathrm{Tt}} / |\pt{t}|^2$ and the fit parameter $k_{\mathrm{T}y}$ is the $y$-component of the transverse momentum of the partons entering the hard scattering. The $k_{\mathrm{T}y}$ parameter needs to be included in the formula, since CCOR only studies distributions on the away side. PHENIX calculates $\rms{\jt{}}$ from a Gaussian fit to the azimuthal angle distribution using the relation
\begin{equation}
\rms{\jt{}} \approx \sqrt{2} \frac{\pt{t} \pt{a}}{\sqrt{\pt{t}^2 + \pt{a}^2}} \sigma_{\mathrm{N}} \,,
\label{eq:phenixjt}
\end{equation}
where $\sigma_{\mathrm{N}}$ is the width of the fitted Gaussian. At the lower collision energies of ISR and RHIC, no evident wide component was observed in the data and thus only one component for $\jt{}$ was extracted by CCOR and PHENIX. This is connected to the current analysis given that especially at the lower energies the high-$\pt{}$ trigger particles are likely to have a high $\mean{z_{\mathrm{t}}}$. PHENIX reported in~\cite{PHENIXjets} that this value is $\mean{z_{\mathrm{t}}} \sim 0.6$. Since ISR had lower collision energy than RHIC, $\mean{z_{\mathrm{t}}}$ can not be lower in the CCOR experiment. In case the trigger particle takes most of the momentum of the leading parton, there is less phase space available for soft gluon radiation during the QCD showering phase. Thus, it appears that the dominant contribution to the particle yield comes from the hadronization part of the fragmentation, and the single component results may be compared to the narrow component results in this analysis.

The PHENIX results are compatible with the ALICE results for bin $0.4 < \xlong < 0.6$ and the CCOR results are close to the ALICE results for bin $0.6 < \xlong < 1.0$. However, a comparison in the same bins is not possible because of the bias $\pt{a}$ selections induce for this analysis.

\section{Conclusions}
\label{sec:conclusions}

A new method to extract two distinct $\jt{}$ components for  a narrow (hadronization) and wide (QCD branching) contribution using two-particle correlations was presented in this work.  The RMS and per trigger yield were obtained for both components. The width of the narrow component shows only a weak dependence on the trigger particle transverse momentum and no difference between $\pp$ and $\pPb$ collisions. The results from this analysis are also qualitatively compatible with the previous ones at lower $\sqrt{s}$, measured by the PHENIX and the CCOR experiments. All of these observations support the universal hadronization expectation. The width of the wide component is found to increase for increasing $\pt{t}$ in all $\xlong$ bins. This can be explained by stronger parton splitting, which is allowed by a larger phase space. A similar argument can be used to explain why the wide component has not been previously observed at the ISR or at RHIC since the larger collision energy at the LHC increases phase space for QCD splittings. As there is no difference in the wide component RMS between $\pp$ and $\pPb$, cold nuclear matter effects do not play a large role in this kinematic regime. \textsc{Pythia}~8 and Herwig~7 simulations describe the widths for both components well, but both simulations overestimate the yield of the narrow component. These measurements could be used to further constrain the parameters in the models to better reproduce the data.

An interesting follow-up study would be to look at the same measurement in heavy-ion collisions. As it is shown that there are no cold nuclear matter effects in $\pPb$ $\jt{}$ distributions, any modifications in the distributions could be attributed to final-state effects, such as partonic energy loss in the quark-gluon plasma. The wide component might be able to discriminate between different jet shape modification mechanisms in $\PbPb$ collisions, like interactions with the plasma~\cite{jetBroadeningAA}, color decoherence effects~\cite{antiangularOrdering}, and changes in relative quark and gluon jet fractions~\cite{jetShapeQGP}.

%%%%% acknowledgements
\newenvironment{acknowledgement}{\relax}{\relax}
\begin{acknowledgement}
\section*{Acknowledgements}
We wish to thank Torbjörn Sjöstrand for his help in defining a di-gluon initial state in \textsc{Pythia}~8.
% Version: 2018-10-16

The ALICE Collaboration would like to thank all its engineers and technicians for their invaluable contributions to the construction of the experiment and the CERN accelerator teams for the outstanding performance of the LHC complex.
The ALICE Collaboration gratefully acknowledges the resources and support provided by all Grid centres and the Worldwide LHC Computing Grid (WLCG) collaboration.
The ALICE Collaboration acknowledges the following funding agencies for their support in building and running the ALICE detector:
A. I. Alikhanyan National Science Laboratory (Yerevan Physics Institute) Foundation (ANSL), State Committee of Science and World Federation of Scientists (WFS), Armenia;
Austrian Academy of Sciences and Nationalstiftung f\"{u}r Forschung, Technologie und Entwicklung, Austria;
Ministry of Communications and High Technologies, National Nuclear Research Center, Azerbaijan;
Conselho Nacional de Desenvolvimento Cient\'{\i}fico e Tecnol\'{o}gico (CNPq), Universidade Federal do Rio Grande do Sul (UFRGS), Financiadora de Estudos e Projetos (Finep) and Funda\c{c}\~{a}o de Amparo \`{a} Pesquisa do Estado de S\~{a}o Paulo (FAPESP), Brazil;
Ministry of Science \& Technology of China (MSTC), National Natural Science Foundation of China (NSFC) and Ministry of Education of China (MOEC) , China;
Ministry of Science and Education, Croatia;
Centro de Aplicaciones Tecnol\'{o}gicas y Desarrollo Nuclear (CEADEN), Cubaenerg\'{\i}a, Cuba;
Ministry of Education, Youth and Sports of the Czech Republic, Czech Republic;
The Danish Council for Independent Research | Natural Sciences, the Carlsberg Foundation and Danish National Research Foundation (DNRF), Denmark;
Helsinki Institute of Physics (HIP), Finland;
Commissariat \`{a} l'Energie Atomique (CEA) and Institut National de Physique Nucl\'{e}aire et de Physique des Particules (IN2P3) and Centre National de la Recherche Scientifique (CNRS), France;
Bundesministerium f\"{u}r Bildung, Wissenschaft, Forschung und Technologie (BMBF) and GSI Helmholtzzentrum f\"{u}r Schwerionenforschung GmbH, Germany;
General Secretariat for Research and Technology, Ministry of Education, Research and Religions, Greece;
National Research, Development and Innovation Office, Hungary;
Department of Atomic Energy Government of India (DAE), Department of Science and Technology, Government of India (DST), University Grants Commission, Government of India (UGC) and Council of Scientific and Industrial Research (CSIR), India;
Indonesian Institute of Science, Indonesia;
Centro Fermi - Museo Storico della Fisica e Centro Studi e Ricerche Enrico Fermi and Istituto Nazionale di Fisica Nucleare (INFN), Italy;
Institute for Innovative Science and Technology , Nagasaki Institute of Applied Science (IIST), Japan Society for the Promotion of Science (JSPS) KAKENHI and Japanese Ministry of Education, Culture, Sports, Science and Technology (MEXT), Japan;
Consejo Nacional de Ciencia (CONACYT) y Tecnolog\'{i}a, through Fondo de Cooperaci\'{o}n Internacional en Ciencia y Tecnolog\'{i}a (FONCICYT) and Direcci\'{o}n General de Asuntos del Personal Academico (DGAPA), Mexico;
Nederlandse Organisatie voor Wetenschappelijk Onderzoek (NWO), Netherlands;
The Research Council of Norway, Norway;
Commission on Science and Technology for Sustainable Development in the South (COMSATS), Pakistan;
Pontificia Universidad Cat\'{o}lica del Per\'{u}, Peru;
Ministry of Science and Higher Education and National Science Centre, Poland;
Korea Institute of Science and Technology Information and National Research Foundation of Korea (NRF), Republic of Korea;
Ministry of Education and Scientific Research, Institute of Atomic Physics and Romanian National Agency for Science, Technology and Innovation, Romania;
Joint Institute for Nuclear Research (JINR), Ministry of Education and Science of the Russian Federation and National Research Centre Kurchatov Institute, Russia;
Ministry of Education, Science, Research and Sport of the Slovak Republic, Slovakia;
National Research Foundation of South Africa, South Africa;
Swedish Research Council (VR) and Knut \& Alice Wallenberg Foundation (KAW), Sweden;
European Organization for Nuclear Research, Switzerland;
National Science and Technology Development Agency (NSDTA), Suranaree University of Technology (SUT) and Office of the Higher Education Commission under NRU project of Thailand, Thailand;
Turkish Atomic Energy Agency (TAEK), Turkey;
National Academy of  Sciences of Ukraine, Ukraine;
Science and Technology Facilities Council (STFC), United Kingdom;
National Science Foundation of the United States of America (NSF) and United States Department of Energy, Office of Nuclear Physics (DOE NP), United States of America.    %%%%%%% done by webmaster team
\end{acknowledgement}

%%%%%%%% Bibliography (In case of using bibtex generate the bbl requested by arXiv)
\bibliographystyle{utphys}   % Remember we use title in the biblio
\bibliography{biblio}

\providecommand{\href}[2]{#2}\begingroup\raggedright\begin{thebibliography}{10}

\bibitem{basicsofpqcd}
Y.~L. Dokshitzer, V.~A. Khoze, A.~H. M{\"u}ller, and S.~I. Troian, {\em {Basics
  of Perturbative QCD}}.
\newblock Editions Fronti{\`e}res, Gif-sur-Yvette, France, 1991.

\bibitem{introPythia81}
T.~Sjöstrand, S.~Mrenna, and P.~Z. Skands, ``{A Brief Introduction to PYTHIA
  8.1},'' \href{http://dx.doi.org/10.1016/j.cpc.2008.01.036}{{\em Comput. Phys.
  Commun.} {\bfseries 178} (2008) 852--867},
\href{http://arxiv.org/abs/0710.3820}{{\ttfamily arXiv:0710.3820 [hep-ph]}}.
%%CITATION = ARXIV:0710.3820;%%.

\bibitem{herwigManual}
M.~Bahr {\em et~al.}, ``{Herwig++ Physics and Manual},''
  \href{http://dx.doi.org/10.1140/epjc/s10052-008-0798-9}{{\em Eur. Phys. J.}
  {\bfseries C58} (2008) 639--707},
\href{http://arxiv.org/abs/0803.0883}{{\ttfamily arXiv:0803.0883 [hep-ph]}}.
%%CITATION = ARXIV:0803.0883;%%.

\bibitem{herwig7releaseNote}
J.~Bellm {\em et~al.}, ``{Herwig 7.0/Herwig++ 3.0 release note},''
  \href{http://dx.doi.org/10.1140/epjc/s10052-016-4018-8}{{\em Eur. Phys. J.}
  {\bfseries C76} no.~4, (2016) 196},
\href{http://arxiv.org/abs/1512.01178}{{\ttfamily arXiv:1512.01178 [hep-ph]}}.
%%CITATION = ARXIV:1512.01178;%%.

\bibitem{firstjtmeasurement}
{\bfseries CCOR} Collaboration, A.~Angelis {\em et~al.}, ``{A Measurement of
  the Transverse Momenta of Partons, and of Jet Fragmentation as a Function of
  $\sqrt{s}$ in $p p$ Collisions},''
\href{http://dx.doi.org/10.1016/0370-2693(80)90572-9}{{\em Phys. Lett.}
  {\bfseries B97} (1980) 163--168}.
%%CITATION = PHLTA,B97,163;%%.

\bibitem{PHENIXjets}
{\bfseries PHENIX} Collaboration, S.~S. Adler {\em et~al.}, ``{Jet properties
  from dihadron correlations in p+p collisions at $\sqrt{s} =$ 200~GeV},''
  \href{http://dx.doi.org/10.1103/PhysRevD.74.072002}{{\em Phys. Rev.}
  {\bfseries D74} (2006) 072002},
\href{http://arxiv.org/abs/hep-ex/0605039}{{\ttfamily arXiv:hep-ex/0605039
  [hep-ex]}}.
%%CITATION = HEP-EX/0605039;%%.

\bibitem{phenixJtPAu}
{\bfseries PHENIX} Collaboration, S.~S. Adler {\em et~al.}, ``{Jet structure
  from dihadron correlations in d+Au collisions at s(NN)**(1/2) = 200-GeV},''
  \href{http://dx.doi.org/10.1103/PhysRevC.73.054903}{{\em Phys. Rev.}
  {\bfseries C73} (2006) 054903},
\href{http://arxiv.org/abs/nucl-ex/0510021}{{\ttfamily arXiv:nucl-ex/0510021
  [nucl-ex]}}.
%%CITATION = NUCL-EX/0510021;%%.

\bibitem{cdfpaper}
{\bfseries CDF} Collaboration, T.~Aaltonen {\em et~al.}, ``{Measurement of the
  $k_T$ Distribution of Particles in Jets Produced in $p\bar{p}$ Collisions at
  $\sqrt{s}=1.96$-TeV},''
  \href{http://dx.doi.org/10.1103/PhysRevLett.102.232002}{{\em Phys. Rev.
  Lett.} {\bfseries 102} (2009) 232002},
\href{http://arxiv.org/abs/0811.2820}{{\ttfamily arXiv:0811.2820 [hep-ex]}}.
%%CITATION = ARXIV:0811.2820;%%.

\bibitem{atlaksenJetit}
{\bfseries ATLAS} Collaboration, A.~Angerami, ``{Measurement of Jets and Jet
  Suppression in $\sqrt{s_{NN}}=2.76$ TeV Lead-Lead Collisions with the ATLAS
  detector at the LHC},''
  \href{http://dx.doi.org/10.1088/0954-3899/38/12/124085}{{\em J. Phys.}
  {\bfseries G38} (2011) 124085},
\href{http://arxiv.org/abs/1108.5191}{{\ttfamily arXiv:1108.5191 [nucl-ex]}}.
%%CITATION = ARXIV:1108.5191;%%.

\bibitem{eventGenerators}
A.~Buckley {\em et~al.}, ``{General-purpose event generators for LHC
  physics},'' \href{http://dx.doi.org/10.1016/j.physrep.2011.03.005}{{\em Phys.
  Rept.} {\bfseries 504} (2011) 145--233},
\href{http://arxiv.org/abs/1101.2599}{{\ttfamily arXiv:1101.2599 [hep-ph]}}.
%%CITATION = ARXIV:1101.2599;%%.

\bibitem{jetBroadeningPpb1}
R.~Baier, Y.~L. Dokshitzer, A.~H. Mueller, S.~Peigne, and D.~Schiff,
  ``{Radiative energy loss and p(T) broadening of high-energy partons in
  nuclei},'' \href{http://dx.doi.org/10.1016/S0550-3213(96)00581-0}{{\em Nucl.
  Phys.} {\bfseries B484} (1997) 265--282},
\href{http://arxiv.org/abs/hep-ph/9608322}{{\ttfamily arXiv:hep-ph/9608322
  [hep-ph]}}.
%%CITATION = HEP-PH/9608322;%%.

\bibitem{introCgc}
L.~D. McLerran and R.~Venugopalan, ``{Computing quark and gluon distribution
  functions for very large nuclei},''
  \href{http://dx.doi.org/10.1103/PhysRevD.49.2233}{{\em Phys. Rev.} {\bfseries
  D49} (1994) 2233--2241},
\href{http://arxiv.org/abs/hep-ph/9309289}{{\ttfamily arXiv:hep-ph/9309289
  [hep-ph]}}.
%%CITATION = HEP-PH/9309289;%%.

\bibitem{eps09}
K.~J. Eskola, H.~Paukkunen, and C.~A. Salgado, ``{EPS09: A New Generation of
  NLO and LO Nuclear Parton Distribution Functions},''
  \href{http://dx.doi.org/10.1088/1126-6708/2009/04/065}{{\em JHEP} {\bfseries
  04} (2009) 065},
\href{http://arxiv.org/abs/0902.4154}{{\ttfamily arXiv:0902.4154 [hep-ph]}}.
%%CITATION = ARXIV:0902.4154;%%.

\bibitem{thorstenBiasProceedings}
T.~Renk, ``{Jet correlations — opportunities and pitfalls},''
  \href{http://dx.doi.org/10.1016/j.nuclphysa.2014.09.066}{{\em Nucl. Phys.}
  {\bfseries A932} (2014) 334--341},
\href{http://arxiv.org/abs/1404.0793}{{\ttfamily arXiv:1404.0793 [hep-ph]}}.
%%CITATION = ARXIV:1404.0793;%%.

\bibitem{Renk:2011wp}
T.~Renk and K.~J. Eskola, ``{Hard dihadron correlations in heavy-ion collisions
  at RHIC and LHC},'' \href{http://dx.doi.org/10.1103/PhysRevC.84.054913}{{\em
  Phys. Rev.} {\bfseries C84} (2011) 054913},
\href{http://arxiv.org/abs/1106.1740}{{\ttfamily arXiv:1106.1740 [hep-ph]}}.
%%CITATION = ARXIV:1106.1740;%%.

\bibitem{aliceDetector}
{\bfseries ALICE} Collaboration, K.~Aamodt {\em et~al.}, ``{The ALICE
  experiment at the CERN LHC},''
\href{http://dx.doi.org/10.1088/1748-0221/3/08/S08002}{{\em JINST} {\bfseries
  3} (2008) S08002}.
%%CITATION = JINST,3,S08002;%%.

\bibitem{alicePerformance}
{\bfseries ALICE} Collaboration, B.~Abelev {\em et~al.}, ``{Performance of the
  ALICE Experiment at the CERN LHC},''
  \href{http://dx.doi.org/10.1142/S0217751X14300440}{{\em Int. J. Mod. Phys.}
  {\bfseries A29} (2014) 1430044},
\href{http://arxiv.org/abs/1402.4476}{{\ttfamily arXiv:1402.4476 [nucl-ex]}}.
%%CITATION = ARXIV:1402.4476;%%.

\bibitem{aliceITS}
{\bfseries ALICE} Collaboration, K.~Aamodt {\em et~al.}, ``{Alignment of the
  ALICE Inner Tracking System with cosmic-ray tracks},''
  \href{http://dx.doi.org/10.1088/1748-0221/5/03/P03003}{{\em JINST} {\bfseries
  5} (2010) P03003},
\href{http://arxiv.org/abs/1001.0502}{{\ttfamily arXiv:1001.0502
  [physics.ins-det]}}.
%%CITATION = ARXIV:1001.0502;%%.

\bibitem{aliceTPC}
J.~Alme {\em et~al.}, ``{The ALICE TPC, a large 3-dimensional tracking device
  with fast readout for ultra-high multiplicity events},''
  \href{http://dx.doi.org/10.1016/j.nima.2010.04.042}{{\em Nucl. Instrum.
  Meth.} {\bfseries A622} (2010) 316--367},
\href{http://arxiv.org/abs/1001.1950}{{\ttfamily arXiv:1001.1950
  [physics.ins-det]}}.
%%CITATION = ARXIV:1001.1950;%%.

\bibitem{aliceBackgroundFluctuation}
{\bfseries ALICE} Collaboration, B.~Abelev {\em et~al.}, ``{Measurement of
  Event Background Fluctuations for Charged Particle Jet Reconstruction in
  Pb-Pb collisions at $\sqrt{s_{NN}} = 2.76$ TeV},''
  \href{http://dx.doi.org/10.1007/JHEP03(2012)053}{{\em JHEP} {\bfseries 03}
  (2012) 053},
\href{http://arxiv.org/abs/1201.2423}{{\ttfamily arXiv:1201.2423 [hep-ex]}}.
%%CITATION = ARXIV:1201.2423;%%.

\bibitem{forwarddetectorsTdr}
{\bfseries ALICE} Collaboration, P.~Cortese {\em et~al.}, ``{ALICE technical
  design report on forward detectors: FMD, T0 and V0},''.
\url{http://cdsweb.cern.ch/record/781854}.
%%CITATION = CERN-LHCC-2004-025;%%.

\bibitem{ALICE:2011ac}
{\bfseries ALICE} Collaboration, B.~Abelev {\em et~al.}, ``{Underlying Event
  measurements in $pp$ collisions at $\sqrt{s}=0.9$ and 7 TeV with the ALICE
  experiment at the LHC},''
  \href{http://dx.doi.org/10.1007/JHEP07(2012)116}{{\em JHEP} {\bfseries 07}
  (2012) 116},
\href{http://arxiv.org/abs/1112.2082}{{\ttfamily arXiv:1112.2082 [hep-ex]}}.
%%CITATION = ARXIV:1112.2082;%%.

\bibitem{hybridExplanation}
{\bfseries ALICE} Collaboration, B.~Abelev {\em et~al.}, ``{Long-range angular
  correlations on the near and away side in $p$-Pb collisions at
  $\sqrt{s_{NN}}=5.02$ TeV},''
  \href{http://dx.doi.org/10.1016/j.physletb.2013.01.012}{{\em Phys. Lett.}
  {\bfseries B719} (2013) 29--41},
\href{http://arxiv.org/abs/1212.2001}{{\ttfamily arXiv:1212.2001 [nucl-ex]}}.
%%CITATION = ARXIV:1212.2001;%%.

\bibitem{introPythia82}
T.~Sjöstrand, S.~Ask, J.~R. Christiansen, R.~Corke, N.~Desai, P.~Ilten,
  S.~Mrenna, S.~Prestel, C.~O. Rasmussen, and P.~Z. Skands, ``{An Introduction
  to PYTHIA 8.2},'' \href{http://dx.doi.org/10.1016/j.cpc.2015.01.024}{{\em
  Comput. Phys. Commun.} {\bfseries 191} (2015) 159--177},
\href{http://arxiv.org/abs/1410.3012}{{\ttfamily arXiv:1410.3012 [hep-ph]}}.
%%CITATION = ARXIV:1410.3012;%%.

\bibitem{pythiaBig}
T.~Sjostrand, S.~Mrenna, and P.~Z. Skands, ``{PYTHIA 6.4 Physics and Manual},''
  \href{http://dx.doi.org/10.1088/1126-6708/2006/05/026}{{\em JHEP} {\bfseries
  05} (2006) 026},
\href{http://arxiv.org/abs/hep-ph/0603175}{{\ttfamily arXiv:hep-ph/0603175
  [hep-ph]}}.
%%CITATION = HEP-PH/0603175;%%.

\bibitem{dpmjet}
S.~Roesler, R.~Engel, and J.~Ranft,
  \href{http://dx.doi.org/10.1007/978-3-642-18211-2_166}{``{The Monte Carlo
  event generator DPMJET-III},''} in {\em {Advanced Monte Carlo for radiation
  physics, particle transport simulation and applications. Proceedings,
  Conference, MC2000, Lisbon, Portugal, October 23-26, 2000}}, pp.~1033--1038.
\newblock 2000.
\newblock \href{http://arxiv.org/abs/hep-ph/0012252}{{\ttfamily
  arXiv:hep-ph/0012252 [hep-ph]}}.
\newblock
\url{http://www-public.slac.stanford.edu/sciDoc/docMeta.aspx?slacPubNumber=SLAC-PUB-8740}.
\newblock
%%CITATION = HEP-PH/0012252;%%.

\bibitem{geant}
R.~Brun, F.~Bruyant, F.~Carminati, S.~Giani, M.~Maire, A.~McPherson,
  G.~Patrick, and L.~Urban, ``{GEANT Detector Description and Simulation
  Tool},''.
\url{http://cds.cern.ch/record/1082634}.
%%CITATION = CERN-W5013;%%.

\bibitem{newPythiaShower}
T.~Sjöstrand and P.~Z. Skands, ``{Transverse-momentum-ordered showers and
  interleaved multiple interactions},''
  \href{http://dx.doi.org/10.1140/epjc/s2004-02084-y}{{\em Eur. Phys. J.}
  {\bfseries C39} (2005) 129--154},
\href{http://arxiv.org/abs/hep-ph/0408302}{{\ttfamily arXiv:hep-ph/0408302
  [hep-ph]}}.
%%CITATION = HEP-PH/0408302;%%.

\bibitem{lundString}
B.~Andersson, G.~Gustafson, G.~Ingelman, and T.~Sj{\"o}strand, ``{Parton
  Fragmentation and String Dynamics},''
\href{http://dx.doi.org/10.1016/0370-1573(83)90080-7}{{\em Phys.Rept.}
  {\bfseries 97} (1983) 31--145}.
%%CITATION = PRPLC,97,31;%%.

\bibitem{spectrumReferencePp}
{\bfseries ALICE} Collaboration, B.~Abelev {\em et~al.}, ``{Energy Dependence
  of the Transverse Momentum Distributions of Charged Particles in pp
  Collisions Measured by ALICE},''
  \href{http://dx.doi.org/10.1140/epjc/s10052-013-2662-9}{{\em Eur. Phys. J.}
  {\bfseries C73} no.~12, (2013) 2662},
\href{http://arxiv.org/abs/1307.1093}{{\ttfamily arXiv:1307.1093 [nucl-ex]}}.
%%CITATION = ARXIV:1307.1093;%%.

\bibitem{spectrumReferencepPb}
{\bfseries ALICE} Collaboration, B.~Abelev {\em et~al.}, ``{Transverse momentum
  dependence of inclusive primary charged-particle production in p-Pb
  collisions at $\sqrt{s_\mathrm{{NN}}}=5.02~\text {TeV}$},''
  \href{http://dx.doi.org/10.1140/epjc/s10052-014-3054-5}{{\em Eur. Phys. J.}
  {\bfseries C74} no.~9, (2014) 3054},
\href{http://arxiv.org/abs/1405.2737}{{\ttfamily arXiv:1405.2737 [nucl-ex]}}.
%%CITATION = ARXIV:1405.2737;%%.

\bibitem{pythia4CTune}
R.~Corke and T.~Sjostrand, ``{Interleaved Parton Showers and Tuning
  Prospects},'' \href{http://dx.doi.org/10.1007/JHEP03(2011)032}{{\em JHEP}
  {\bfseries 03} (2011) 032},
\href{http://arxiv.org/abs/1011.1759}{{\ttfamily arXiv:1011.1759 [hep-ph]}}.
%%CITATION = ARXIV:1011.1759;%%.

\bibitem{pythiaMonashTune}
P.~Skands, S.~Carrazza, and J.~Rojo, ``{Tuning PYTHIA 8.1: the Monash 2013
  Tune},'' \href{http://dx.doi.org/10.1140/epjc/s10052-014-3024-y}{{\em Eur.
  Phys. J.} {\bfseries C74} no.~8, (2014) 3024},
\href{http://arxiv.org/abs/1404.5630}{{\ttfamily arXiv:1404.5630 [hep-ph]}}.
%%CITATION = ARXIV:1404.5630;%%.

\bibitem{jetBroadeningAA}
C.~A. Salgado and U.~A. Wiedemann, ``{Medium modification of jet shapes and jet
  multiplicities},''
  \href{http://dx.doi.org/10.1103/PhysRevLett.93.042301}{{\em Phys. Rev. Lett.}
  {\bfseries 93} (2004) 042301},
\href{http://arxiv.org/abs/hep-ph/0310079}{{\ttfamily arXiv:hep-ph/0310079
  [hep-ph]}}.
%%CITATION = HEP-PH/0310079;%%.

\bibitem{antiangularOrdering}
Y.~Mehtar-Tani, C.~A. Salgado, and K.~Tywoniuk, ``{Anti-angular ordering of
  gluon radiation in QCD media},''
  \href{http://dx.doi.org/10.1103/PhysRevLett.106.122002}{{\em Phys. Rev.
  Lett.} {\bfseries 106} (2011) 122002},
\href{http://arxiv.org/abs/1009.2965}{{\ttfamily arXiv:1009.2965 [hep-ph]}}.
%%CITATION = ARXIV:1009.2965;%%.

\bibitem{jetShapeQGP}
Y.-T. Chien and I.~Vitev, ``{Towards the understanding of jet shapes and cross
  sections in heavy ion collisions using soft-collinear effective theory},''
  \href{http://dx.doi.org/10.1007/JHEP05(2016)023}{{\em JHEP} {\bfseries 05}
  (2016) 023},
\href{http://arxiv.org/abs/1509.07257}{{\ttfamily arXiv:1509.07257 [hep-ph]}}.
%%CITATION = ARXIV:1509.07257;%%.

\end{thebibliography}\endgroup
%\input {bibliography.tex}  

%%%%%%%%% appendix with author list
\newpage
\appendix
\section{The ALICE Collaboration}
\label{app:collab}
% Collaboration: CERN-LHC-ALICE
% Generation Date is 2018-Oct-16

% How to use:
%%%%%%%%% appendix with author list
%\appendix
%\section{The ALICE Collaboration}
%\label{app:collab}
%\input{Alice_Authorslist_XXXX-Axx-XX.tex}
\begingroup
\small
\begin{flushleft}
S.~Acharya\Irefn{org140}\And 
F.T.-.~Acosta\Irefn{org20}\And 
D.~Adamov\'{a}\Irefn{org93}\And 
S.P.~Adhya\Irefn{org140}\And 
A.~Adler\Irefn{org74}\And 
J.~Adolfsson\Irefn{org80}\And 
M.M.~Aggarwal\Irefn{org98}\And 
G.~Aglieri Rinella\Irefn{org34}\And 
M.~Agnello\Irefn{org31}\And 
Z.~Ahammed\Irefn{org140}\And 
S.~Ahmad\Irefn{org17}
% ...
\And
\renewcommand\labelenumi{\textsuperscript{\theenumi}~}

\section*{Affiliation notes}
\renewcommand\theenumi{\roman{enumi}}
\begin{Authlist}
\item \Adef{org*}Deceased
\item \Adef{orgI}Dipartimento DET del Politecnico di Torino, Turin, Italy
\item \Adef{orgII}M.V. Lomonosov Moscow State University, D.V. Skobeltsyn Institute of Nuclear, Physics, Moscow, Russia
\item \Adef{orgIII}Department of Applied Physics, Aligarh Muslim University, Aligarh, India
\item \Adef{orgIV}Institute of Theoretical Physics, University of Wroclaw, Poland
\end{Authlist}

\section*{Collaboration Institutes}
\renewcommand\theenumi{\arabic{enumi}~}
\begin{Authlist}
\item \Idef{org1}A.I. Alikhanyan National Science Laboratory (Yerevan Physics Institute) Foundation, Yerevan, Armenia
\item \Idef{org2}Bogolyubov Institute for Theoretical Physics, National Academy of Sciences of Ukraine, Kiev, Ukraine
\item \Idef{org3}Bose Institute, Department of Physics  and Centre for Astroparticle Physics and Space Science (CAPSS), Kolkata, India
\item \Idef{org4}Budker Institute for Nuclear Physics, Novosibirsk, Russia
\item \Idef{org5}California Polytechnic State University, San Luis Obispo, California, United States
\item \Idef{org6}Central China Normal University, Wuhan, China
\item \Idef{org7}Centre de Calcul de l'IN2P3, Villeurbanne, Lyon, France
\item \Idef{org8}Centro de Aplicaciones Tecnol\'{o}gicas y Desarrollo Nuclear (CEADEN), Havana, Cuba
\item \Idef{org9}Centro de Investigaci\'{o}n y de Estudios Avanzados (CINVESTAV), Mexico City and M\'{e}rida, Mexico
\item \Idef{org10}Centro Fermi - Museo Storico della Fisica e Centro Studi e Ricerche ``Enrico Fermi', Rome, Italy
\item \Idef{org11}Chicago State University, Chicago, Illinois, United States
\item \Idef{org12}China Institute of Atomic Energy, Beijing, China
\item \Idef{org13}Chonbuk National University, Jeonju, Republic of Korea
\item \Idef{org14}Comenius University Bratislava, Faculty of Mathematics, Physics and Informatics, Bratislava, Slovakia
\item \Idef{org15}COMSATS Institute of Information Technology (CIIT), Islamabad, Pakistan
\item \Idef{org16}Creighton University, Omaha, Nebraska, United States
\item \Idef{org17}Department of Physics, Aligarh Muslim University, Aligarh, India
\item \Idef{org18}Department of Physics, Pusan National University, Pusan, Republic of Korea
\item \Idef{org19}Department of Physics, Sejong University, Seoul, Republic of Korea
\item \Idef{org20}Department of Physics, University of California, Berkeley, California, United States
\item \Idef{org21}Department of Physics, University of Oslo, Oslo, Norway
\item \Idef{org22}Department of Physics and Technology, University of Bergen, Bergen, Norway
\item \Idef{org23}Dipartimento di Fisica dell'Universit\`{a} 'La Sapienza' and Sezione INFN, Rome, Italy
\item \Idef{org24}Dipartimento di Fisica dell'Universit\`{a} and Sezione INFN, Cagliari, Italy
\item \Idef{org25}Dipartimento di Fisica dell'Universit\`{a} and Sezione INFN, Trieste, Italy
\item \Idef{org26}Dipartimento di Fisica dell'Universit\`{a} and Sezione INFN, Turin, Italy
\item \Idef{org27}Dipartimento di Fisica e Astronomia dell'Universit\`{a} and Sezione INFN, Bologna, Italy
\item \Idef{org28}Dipartimento di Fisica e Astronomia dell'Universit\`{a} and Sezione INFN, Catania, Italy
\item \Idef{org29}Dipartimento di Fisica e Astronomia dell'Universit\`{a} and Sezione INFN, Padova, Italy
\item \Idef{org30}Dipartimento di Fisica `E.R.~Caianiello' dell'Universit\`{a} and Gruppo Collegato INFN, Salerno, Italy
\item \Idef{org31}Dipartimento DISAT del Politecnico and Sezione INFN, Turin, Italy
\item \Idef{org32}Dipartimento di Scienze e Innovazione Tecnologica dell'Universit\`{a} del Piemonte Orientale and INFN Sezione di Torino, Alessandria, Italy
\item \Idef{org33}Dipartimento Interateneo di Fisica `M.~Merlin' and Sezione INFN, Bari, Italy
\item \Idef{org34}European Organization for Nuclear Research (CERN), Geneva, Switzerland
\item \Idef{org35}Faculty of Electrical Engineering, Mechanical Engineering and Naval Architecture, University of Split, Split, Croatia
\item \Idef{org36}Faculty of Engineering and Science, Western Norway University of Applied Sciences, Bergen, Norway
\item \Idef{org37}Faculty of Nuclear Sciences and Physical Engineering, Czech Technical University in Prague, Prague, Czech Republic
\item \Idef{org38}Faculty of Science, P.J.~\v{S}af\'{a}rik University, Ko\v{s}ice, Slovakia
\item \Idef{org39}Frankfurt Institute for Advanced Studies, Johann Wolfgang Goethe-Universit\"{a}t Frankfurt, Frankfurt, Germany
\item \Idef{org40}Gangneung-Wonju National University, Gangneung, Republic of Korea
\item \Idef{org41}Gauhati University, Department of Physics, Guwahati, India
\item \Idef{org42}Helmholtz-Institut f\"{u}r Strahlen- und Kernphysik, Rheinische Friedrich-Wilhelms-Universit\"{a}t Bonn, Bonn, Germany
\item \Idef{org43}Helsinki Institute of Physics (HIP), Helsinki, Finland
\item \Idef{org44}High Energy Physics Group,  Universidad Aut\'{o}noma de Puebla, Puebla, Mexico
\item \Idef{org45}Hiroshima University, Hiroshima, Japan
\item \Idef{org46}Hochschule Worms, Zentrum  f\"{u}r Technologietransfer und Telekommunikation (ZTT), Worms, Germany
\item \Idef{org47}Horia Hulubei National Institute of Physics and Nuclear Engineering, Bucharest, Romania
\item \Idef{org48}Indian Institute of Technology Bombay (IIT), Mumbai, India
\item \Idef{org49}Indian Institute of Technology Indore, Indore, India
\item \Idef{org50}Indonesian Institute of Sciences, Jakarta, Indonesia
\item \Idef{org51}INFN, Laboratori Nazionali di Frascati, Frascati, Italy
\item \Idef{org52}INFN, Sezione di Bari, Bari, Italy
\item \Idef{org53}INFN, Sezione di Bologna, Bologna, Italy
\item \Idef{org54}INFN, Sezione di Cagliari, Cagliari, Italy
\item \Idef{org55}INFN, Sezione di Catania, Catania, Italy
\item \Idef{org56}INFN, Sezione di Padova, Padova, Italy
\item \Idef{org57}INFN, Sezione di Roma, Rome, Italy
\item \Idef{org58}INFN, Sezione di Torino, Turin, Italy
\item \Idef{org59}INFN, Sezione di Trieste, Trieste, Italy
\item \Idef{org60}Inha University, Incheon, Republic of Korea
\item \Idef{org61}Institut de Physique Nucl\'{e}aire d'Orsay (IPNO), Institut National de Physique Nucl\'{e}aire et de Physique des Particules (IN2P3/CNRS), Universit\'{e} de Paris-Sud, Universit\'{e} Paris-Saclay, Orsay, France
\item \Idef{org62}Institute for Nuclear Research, Academy of Sciences, Moscow, Russia
\item \Idef{org63}Institute for Subatomic Physics, Utrecht University/Nikhef, Utrecht, Netherlands
\item \Idef{org64}Institute for Theoretical and Experimental Physics, Moscow, Russia
\item \Idef{org65}Institute of Experimental Physics, Slovak Academy of Sciences, Ko\v{s}ice, Slovakia
\item \Idef{org66}Institute of Physics, Homi Bhabha National Institute, Bhubaneswar, India
\item \Idef{org67}Institute of Physics of the Czech Academy of Sciences, Prague, Czech Republic
\item \Idef{org68}Institute of Space Science (ISS), Bucharest, Romania
\item \Idef{org69}Institut f\"{u}r Kernphysik, Johann Wolfgang Goethe-Universit\"{a}t Frankfurt, Frankfurt, Germany
\item \Idef{org70}Instituto de Ciencias Nucleares, Universidad Nacional Aut\'{o}noma de M\'{e}xico, Mexico City, Mexico
\item \Idef{org71}Instituto de F\'{i}sica, Universidade Federal do Rio Grande do Sul (UFRGS), Porto Alegre, Brazil
\item \Idef{org72}Instituto de F\'{\i}sica, Universidad Nacional Aut\'{o}noma de M\'{e}xico, Mexico City, Mexico
\item \Idef{org73}iThemba LABS, National Research Foundation, Somerset West, South Africa
\item \Idef{org74}Johann-Wolfgang-Goethe Universit\"{a}t Frankfurt Institut f\"{u}r Informatik, Fachbereich Informatik und Mathematik, Frankfurt, Germany
\item \Idef{org75}Joint Institute for Nuclear Research (JINR), Dubna, Russia
\item \Idef{org76}Korea Institute of Science and Technology Information, Daejeon, Republic of Korea
\item \Idef{org77}KTO Karatay University, Konya, Turkey
\item \Idef{org78}Laboratoire de Physique Subatomique et de Cosmologie, Universit\'{e} Grenoble-Alpes, CNRS-IN2P3, Grenoble, France
\item \Idef{org79}Lawrence Berkeley National Laboratory, Berkeley, California, United States
\item \Idef{org80}Lund University Department of Physics, Division of Particle Physics, Lund, Sweden
\item \Idef{org81}Nagasaki Institute of Applied Science, Nagasaki, Japan
\item \Idef{org82}Nara Women{'}s University (NWU), Nara, Japan
\item \Idef{org83}National and Kapodistrian University of Athens, School of Science, Department of Physics , Athens, Greece
\item \Idef{org84}National Centre for Nuclear Research, Warsaw, Poland
\item \Idef{org85}National Institute of Science Education and Research, Homi Bhabha National Institute, Jatni, India
\item \Idef{org86}National Nuclear Research Center, Baku, Azerbaijan
\item \Idef{org87}National Research Centre Kurchatov Institute, Moscow, Russia
\item \Idef{org88}Niels Bohr Institute, University of Copenhagen, Copenhagen, Denmark
\item \Idef{org89}Nikhef, National institute for subatomic physics, Amsterdam, Netherlands
\item \Idef{org90}NRC Kurchatov Institute IHEP, Protvino, Russia
\item \Idef{org91}NRNU Moscow Engineering Physics Institute, Moscow, Russia
\item \Idef{org92}Nuclear Physics Group, STFC Daresbury Laboratory, Daresbury, United Kingdom
\item \Idef{org93}Nuclear Physics Institute of the Czech Academy of Sciences, \v{R}e\v{z} u Prahy, Czech Republic
\item \Idef{org94}Oak Ridge National Laboratory, Oak Ridge, Tennessee, United States
\item \Idef{org95}Ohio State University, Columbus, Ohio, United States
\item \Idef{org96}Petersburg Nuclear Physics Institute, Gatchina, Russia
\item \Idef{org97}Physics department, Faculty of science, University of Zagreb, Zagreb, Croatia
\item \Idef{org98}Physics Department, Panjab University, Chandigarh, India
\item \Idef{org99}Physics Department, University of Jammu, Jammu, India
\item \Idef{org100}Physics Department, University of Rajasthan, Jaipur, India
\item \Idef{org101}Physikalisches Institut, Eberhard-Karls-Universit\"{a}t T\"{u}bingen, T\"{u}bingen, Germany
\item \Idef{org102}Physikalisches Institut, Ruprecht-Karls-Universit\"{a}t Heidelberg, Heidelberg, Germany
\item \Idef{org103}Physik Department, Technische Universit\"{a}t M\"{u}nchen, Munich, Germany
\item \Idef{org104}Research Division and ExtreMe Matter Institute EMMI, GSI Helmholtzzentrum f\"ur Schwerionenforschung GmbH, Darmstadt, Germany
\item \Idef{org105}Rudjer Bo\v{s}kovi\'{c} Institute, Zagreb, Croatia
\item \Idef{org106}Russian Federal Nuclear Center (VNIIEF), Sarov, Russia
\item \Idef{org107}Saha Institute of Nuclear Physics, Homi Bhabha National Institute, Kolkata, India
\item \Idef{org108}School of Physics and Astronomy, University of Birmingham, Birmingham, United Kingdom
\item \Idef{org109}Secci\'{o}n F\'{\i}sica, Departamento de Ciencias, Pontificia Universidad Cat\'{o}lica del Per\'{u}, Lima, Peru
\item \Idef{org110}Shanghai Institute of Applied Physics, Shanghai, China
\item \Idef{org111}St. Petersburg State University, St. Petersburg, Russia
\item \Idef{org112}Stefan Meyer Institut f\"{u}r Subatomare Physik (SMI), Vienna, Austria
\item \Idef{org113}SUBATECH, IMT Atlantique, Universit\'{e} de Nantes, CNRS-IN2P3, Nantes, France
\item \Idef{org114}Suranaree University of Technology, Nakhon Ratchasima, Thailand
\item \Idef{org115}Technical University of Ko\v{s}ice, Ko\v{s}ice, Slovakia
\item \Idef{org116}Technische Universit\"{a}t M\"{u}nchen, Excellence Cluster 'Universe', Munich, Germany
\item \Idef{org117}The Henryk Niewodniczanski Institute of Nuclear Physics, Polish Academy of Sciences, Cracow, Poland
\item \Idef{org118}The University of Texas at Austin, Austin, Texas, United States
\item \Idef{org119}Universidad Aut\'{o}noma de Sinaloa, Culiac\'{a}n, Mexico
\item \Idef{org120}Universidade de S\~{a}o Paulo (USP), S\~{a}o Paulo, Brazil
\item \Idef{org121}Universidade Estadual de Campinas (UNICAMP), Campinas, Brazil
\item \Idef{org122}Universidade Federal do ABC, Santo Andre, Brazil
\item \Idef{org123}University College of Southeast Norway, Tonsberg, Norway
\item \Idef{org124}University of Cape Town, Cape Town, South Africa
\item \Idef{org125}University of Houston, Houston, Texas, United States
\item \Idef{org126}University of Jyv\"{a}skyl\"{a}, Jyv\"{a}skyl\"{a}, Finland
\item \Idef{org127}University of Liverpool, Liverpool, United Kingdom
\item \Idef{org128}University of Science and Techonology of China, Hefei, China
\item \Idef{org129}University of Tennessee, Knoxville, Tennessee, United States
\item \Idef{org130}University of the Witwatersrand, Johannesburg, South Africa
\item \Idef{org131}University of Tokyo, Tokyo, Japan
\item \Idef{org132}University of Tsukuba, Tsukuba, Japan
\item \Idef{org133}Universit\'{e} Clermont Auvergne, CNRS/IN2P3, LPC, Clermont-Ferrand, France
\item \Idef{org134}Universit\'{e} de Lyon, Universit\'{e} Lyon 1, CNRS/IN2P3, IPN-Lyon, Villeurbanne, Lyon, France
\item \Idef{org135}Universit\'{e} de Strasbourg, CNRS, IPHC UMR 7178, F-67000 Strasbourg, France, Strasbourg, France
\item \Idef{org136} Universit\'{e} Paris-Saclay Centre d¿\'Etudes de Saclay (CEA), IRFU, Department de Physique Nucl\'{e}aire (DPhN), Saclay, France
\item \Idef{org137}Universit\`{a} degli Studi di Foggia, Foggia, Italy
\item \Idef{org138}Universit\`{a} degli Studi di Pavia, Pavia, Italy
\item \Idef{org139}Universit\`{a} di Brescia, Brescia, Italy
\item \Idef{org140}Variable Energy Cyclotron Centre, Homi Bhabha National Institute, Kolkata, India
\item \Idef{org141}Warsaw University of Technology, Warsaw, Poland
\item \Idef{org142}Wayne State University, Detroit, Michigan, United States
\item \Idef{org143}Westf\"{a}lische Wilhelms-Universit\"{a}t M\"{u}nster, Institut f\"{u}r Kernphysik, M\"{u}nster, Germany
\item \Idef{org144}Wigner Research Centre for Physics, Hungarian Academy of Sciences, Budapest, Hungary
\item \Idef{org145}Yale University, New Haven, Connecticut, United States
\item \Idef{org146}Yonsei University, Seoul, Republic of Korea
\end{Authlist}
\endgroup
  %%%%%%% done by webmaster team
\end{document}